\documentclass[conf]{new-aiaa}
\usepackage{booktabs}
\usepackage{algorithm}
\usepackage{algorithmic}
\usepackage{new-mysymb}

\newcommand\hvy{\hat{\mathbf{y}}}
\newcommand\hvY{\hat{\mathbf{Y}}}
\newcommand\tr{\mathrm{tr}}
\newcommand\koop{\cK_t}

\newcommand\tvL{\Tilde{\vL}}

\newtheorem{theorem}{Theorem}
\title{Learning Networked Dynamical System Models with \\ Weak Form and Graph Neural Networks}

\author{Yin Yu\footnote{Graduate Student Research Assistant, Department of Aerospace Engineering, Student Member AIAA, yzy5368@psu.edu}}
\author{Daning Huang\footnote{Assistant Professor, Department of Aerospace Engineering, Member AIAA, daning@psu.edu}}
\author{Seho Park\footnote{Graduate Student Research Assistant, Department of Mechanical Engineering, sehopark@psu.edu}}
\author{Herschel C. Pangborn\footnote{Assistant Professor, Department of Mechanical Engineering, Member, AIAA, hcpangborn@psu.edu}}
\affil{The Pennsylvania State University, University Park, PA, 16802}

\begin{document}
\maketitle
\doublespacing

\begin{abstract}
    This paper presents a sequence of two approaches for the data-driven control-oriented modeling of networked systems, i.e., the systems that involve many interacting dynamical components. First, a novel deep learning approach named the weak Latent Dynamics Model (wLDM) is developed for learning generic nonlinear dynamics with control.  Leveraging the weak form, the wLDM enables more numerically stable and computationally efficient training as well as more accurate prediction, when compared to conventional methods such as neural ordinary differential equations. Building upon the wLDM framework, we propose the weak Graph Koopman Bilinear Form (wGKBF) model, which integrates geometric deep learning and Koopman theory to learn latent space dynamics for networked systems, especially for the challenging cases having multiple timescales. The effectiveness of the wLDM framework and wGKBF model are demonstrated on three example systems of increasing complexity - a controlled double pendulum, the stiff Brusselator dynamics, and an electrified aircraft energy system. These numerical examples show that the wLDM and wGKBF achieve superior predictive accuracy and training efficiency as compared to baseline models. Parametric studies provide insights into the effects of hyperparameters in the weak form. The proposed framework shows the capability to efficiently capture control-dependent dynamics in these systems, including stiff dynamics and multi-physics interactions, offering a promising direction for learning control-oriented models of complex networked systems.
\end{abstract}

\section*{Nomenclature}

{\renewcommand\arraystretch{1.0}
\textit{Roman Symbols}
\noindent\begin{longtable*}{@{}l @{\quad=\quad} l@{}}
$a$, $b$                    & Time interval bounds of test functions \\
$A$, $B$                    & Brusselator system parameters \\
$\vA$, $\vB_i$              & System matrices of bilinear dynamical systems\\
$\vC$, $\vD$                & Weight matrices of integration approximation \\
$\cD$                       & Dataset \\
$\cE$                       & Edges of a graph \\
$\vf$                       & System dynamics \\
$\vf_E$, $\vf_P$, $\vf_D$   & Encoder, processor, and decoder of neural network \\
$\vg$                       & Observation function \\
$\cG$                       & Graph \\
$\vh$, $\vH$                & Node feature vectors and their collection \\
% $H$                         & Hamiltonian function \\
$J$                         & Optimization objective \\
$\koop$                     & Koopman operator\\
$\cL$                       & Loss function \\
$L$                         & Number of steps in a trajectory \\
% $L_i$                       & Double pendulum length \\
% $\cL_{\vf}$, $\cL_{\vg}$    & Lie derivatives \\
$\vL$                       & Laplacian matrix of a graph \\
% $M_i$,                      & Double pendulum mass \\
% $M_{ag}$,$M_{up}$           & Message passing \\
% $\vM$,                      & Diagonal degree matrix \\
% $n_x$,$n_u$,$n_y$,$n_\theta$,$n_D$ & Dimension of corresponding quantities \\
$N$                         & Time horizon, i.e., number of steps in a time interval of a test function \\
$N_E$, $N_P$, $N_D$         & Number of layers for encoder, processor, and decoder \\
$\cN_I$                     & Neighbors of node $I$ \\
$n_I,m_I$                   & Dimensions of states and inputs at node $I$ \\
$n_N$                       & Number of nodes \\
$n_T$                       & Number of trajectories \\
$P^{(\alpha,\beta)}$        & Jacobi polynomial \\
$S$                         & Latent dimension \\
$\tilde{S}$                 & Latent dimension of a node \\
% $\vS$                       & Adjacency matrix \\
% $S_R$                       & Stiffness ratio \\
% $T$                         & Prediction horizon \\
$T_k$                       & The $k$th order Chebyshev polynomial \\
$t$                         & Time \\
$\Delta t$                  & Time step \\
$\vu$, $\vU$                & Control input vector and its collection \\
$\cV$                       & Nodes of a graph \\
$\mathbf{v}$, $v_i$         & Numerical integration weights and the $i$th element \\
$\vw$, $\vW$                & Latent representation vector and its collection \\
% $\cW$                       & Graph edge weights \\
$\vx$, $\vX$                & State vector and its collection \\
$\vy$, $\vY$                & System observation and its collection \\
$\hat{\vy}$, $\hat{\vY}$    & Measured observation and its collection
% $\hat{y}_i$,$\hvy$, $\hvY$  & Noisy system observations \\
% $\tvy$, $\tvY$              & Predicted system observations \\
\end{longtable*}
\textit{Greek Symbols}
\noindent\begin{longtable*}{@{}l @{\quad=\quad} l@{}}
$\alpha$, $\beta$           & Jacobi polynomial parameters \\
$\gamma$                    & Penalty parameter \\
% $\delta$, $\theta_i$        & Double pendulum angles \\
$\delta_{ij}$               & Kronecker delta \\
% $\eta(\cdot)$               & Integrable function \\
$\vtQ$                      & Trainable parameters of the model \\
% $\vtL$, $\lambda$           & Eigenvalues \\ 
$\vtl$                      & Adjoint state vector \\
$\rho$                      & Weight function for the weak form test functions\\
$\vts$                      & Activation function \\
% $\tau$                      & Interval \\
% $\vtF$, $\vtY$              & Matrices of right and left eigenvectors\\
% $\vtvf, \varphi_i$          & Eigenfunctions, $i$th eigenfunction \\
% $\vtf(\cdot)$,$\vty(\cdot)$ & Mapping for lifted coordinates\\
$\phi$                      & Test functions
\end{longtable*}
\textit{Others}
\noindent\begin{longtable*}{@{}l @{\quad=\quad} l@{}}
$\hat{\Box}$                & Measured observation for training \\
$\Box_I$                    & Index for nodes of a graph \\
$\Box_i$                    & Index for time or element \\
$n_\Box$                    & Dimension of the quantity $\Box$
% $\circ$                     & Function composition \\
\end{longtable*}
\textit{Acronyms}
\noindent\begin{longtable*}{@{}l @{\quad=\quad} l@{}}
BDF                         & Backward Differentiation Formula \\
biDMDc                      & bilinear Dynamic Mode Decomposition with control \\
FCNN                        & Fully-Connected Neural Network \\
GCL                         & Graph Convolution Layer \\
GNN                         & Graph Neural Network \\
IO                          & Integral Order \\
KBF                         & Koopman Bilinear Form \\
LSTM                        & Long Short-Term Memory \\
MP                          & Message Passing \\
MPC                         & Model Predictive Control \\
NODE                        & Neural Ordinary Differential Equation \\
NRMSE                       & Normalized Root Mean Squared Error \\
ODE                         & Ordinary Differential Equation \\
PO                          & Polynomial Order \\
PReLU                       & Parametric Rectified Linear Unit \\
wLDM                        & weak Latent Dynamics Model \\
wGKBF                       & weak Graph Koopman Bilinear Form \\
SINDy                       & Sparse Identification of Nonlinear Dynamics
\end{longtable*}}

% -----------------------------------------------
\section{Introduction}
% -----------------------------------------------
Powertrain electrification is a megatrend in vehicle design, leading to enhanced performance, new paradigms in safety and sustainability, and decreased lifetime costs \cite{Boglietti2009,Roberts2014b,Karimi2007,Gerstler2008,Park2021}. Electrified vehicle energy systems are characterized by complex nonlinear dynamics, having tightly coupled interactions among system components across multiple energy domains (e.g., electrical, thermal, mechanical).
This is typical of many networked systems, which consist of many components whose dynamics may affect each other.
Networked systems usually demonstrate two unique features: sparsity and heterogeneity \cite{Newman2018}.
Sparsity refers to a coupling structure in which each component only interacts with a small number of other components. This is a feature that can be leveraged to simplify the modeling and control of networked systems.
Heterogeneity implies that the components may be governed by different dynamics and interact with a different number of other components.  Importantly, the dynamics across all the components may span a wide range of timescales, thus producing a stiff system as a whole.  Heterogeneity, together with the potential for stiffness, typically pose a challenge to the modeling and control of networked systems.

Control methods for nonlinear systems have been explored and demonstrated extensively in the literature \cite{Slotine1991,Khalil2002}.  A notable example for the networked systems is hierarchical Model Predictive Control (MPC) using graph-based models \cite{Aksland2022,Koeln2019}, which is particularly effective at coordinating decision-making across energy systems with multiple timescales.
The performance and robustness of such methods are highly sensitive to model error, yet an accurate model is usually difficult and time-consuming to obtain, especially for complex systems such as the electrified vehicle energy systems.
For example, even if physics-based equations are employed to model the energy system, there may still be unknown functional terms and parameters, e.g., electrical efficiencies and heat transfer coefficients, that need to be estimated empirically.
In addition, the physics-based model may have nonlinear terms that are inconvenient for controller design \cite{Park2023}.

% Furthermore, these control methods can become increasingly computationally expensive as the state dimension increases, especially in the presence of nonlinearities \cite{Park2023}, making them infeasible for real-time applications.

This paper aims to develop methods for creating data-driven models of networked systems that are suitable for use in model-based control design. The key questions are twofold: (1) What is the appropriate form for such a control-oriented model, and (2) how to identify the model from data?

\paragraph{Choice of Model Form}
A generic answer to the first question is that a control-oriented model should be sufficiently compact, i.e., low-dimensional, for computational efficiency and hardware deployment. Ideally the model should possess a  structure that a control method can leverage \cite{Slotine1991,Khalil2002}, with classical examples being transfer functions and linear time-invariant state space models.  Relaxing the linearity requirement, another example is control-affine models that are only linear in the control inputs \cite{Khalil2002}.
A control-oriented modeling approach undergoing recent revival is the bilinear model \cite{Bruder2021}, which has a special, mildly nonlinear control-affine form whose autonomous component is linear in states and control component is bilinear in both states and control.  One advantage of bilinear models over generic control-affine models is the availability of various specialized, computationally efficient control methods \cite{Kanai2023,Lian2023}.
A second advantage is the theoretical foundation in
Koopman theory \cite{Budisic2012, Surana2016, Mauroy2016, mauroy2020koopman}, particularly the Koopman Bilinear Form (KBF) \cite{Goldschmidt2021, Goswami2022, Jiang2024}, which provides rigorous conditions for the existence of bilinear models for the representation of a broad range of nonlinear dynamics \cite{abraham2019active, Rosenfeld2021, schulze2022data}.
In Koopman theory, nonlinear dynamics are lifted to a higher-dimensional space where the dynamics are governed by a linear operator.  Notable data-driven methods include extended DMD with control (eDMDc) \cite{Proctor2016} and bilinear DMD with control (biDMDc) \cite{Goldschmidt2021}. These conventional Koopman methods typically use a library of feature functions, e.g., polynomials, to construct the mapping between physical and Koopman states.  Such library-based mapping poses significant challenges, particularly in maintaining predictive accuracy and generalizability, as well as scalability to high-dimensional physical states. Recent approaches leverage deep neural networks to learn this mapping \cite{Li2017, Huang2023koopman}, and the resulting models show improvements over conventional Koopman methods that use library-based mapping in model compactness, predictive accuracy, and control performance \cite{Folkestad2022, shi2022deep}.

Despite the success of data-driven methods for identifying Koopman operators, the extension of KBF to networked systems remains underexplored due to the challenge in the data representation of networked systems.
It is natural to use graphs to describe the networked systems, where the nodes represent the components and the edges represent the interactions.
However, traditional machine learning algorithms, such as convolutional neural networks, are defined for structured data, e.g., vectors and matrices, and struggle with unstructured data, e.g., those defined on graphs \cite{Battaglia2018}.
The unstructured data do not conform to a tabular structure such as matrices, and conventional tools such as linear algebra cannot be directly applied.
Graph Neural Networks (GNNs) \cite{Scarselli2009,Hamilton2020} have emerged as a powerful tool for modeling data defined on graphs, learning from both node-level and edge-level features to capture local and global patterns. They handle unstructured data that is common in real-world applications.  In particular, incorporating domain knowledge and system properties into graph structures has been found to better characterize the underlying dynamics and interactions, leading to more accurate and efficient dynamical models \cite{Yu2024c,Yu2024}.
GNN-based models have been successfully applied to complex physical simulations, such as fluid dynamics \cite{Yu2022,Massegur2024}, weather systems \cite{Li2023,Yu2024b}, robotic control systems \cite{Sanchez2018, Li2019}, power systems \cite{Karimi2021,Yu2024}, etc.
Hence, the extension of KBF with GNNs is worth exploring, as it holds potential to produce a computationally efficient control-oriented model that encodes interactions among the components of a complex networked system while efficiently handling sparse connectivity and heterogeneous dynamics.

% While recent works incorporate graph-based representations into the  framework \cite{Fujii2019}, they do not explicitly consider the bilinear form of the Koopman operator or the effect of control inputs, which are essential for modeling and control of nonlinear systems \cite{Surana2016}.

\paragraph{Learning Dynamics from Data}
After selecting an appropriate model form, the next step is to learn the system dynamics from available data. % The learning process involves identifying the underlying relationships and dependencies among the system components and estimating the model parameters that best capture the observed behavior.
The algorithms for learning dynamics are broadly categorized into discrete-time and continuous-time formulations. Discrete-time methods, such as autoregressive models \cite{Lutkepohl2005} and recurrent neural networks \cite{Hochreiter1997long}, map the system state from one time step to the next. These methods are easy to implement but may have highly nonlinear model forms that are inconvenient for controller design.  Furthermore, it is either difficult or computationally inefficient to adapt the model to different time step sizes. This limitation poses significant challenges in various control scenarios, such as hierarchical MPC, where the ability to handle multiple time step sizes is essential \cite{Yu2023}. In contrast, continuous-time methods learn a model to map the states to their time derivatives, offering a more natural and often more compact representation of underlying dynamical processes, especially when the underlying physics are governed by ordinary differential equations (ODEs).  In control applications, the ODE may be discretized differently to maximize the computational efficiency to resolve the dynamics at different timescales.

One major category of continuous-time methods is the ``derivative-based'' formulation, which directly fits a model to match the time derivative of the states \cite{Brunton2022}. This formulation usually results in a linear or nonlinear constraint-free regression problem.
A representative and widely used method in this category is Sparse Identification of Nonlinear Dynamics (SINDy) \cite{Brunton2016}. SINDy builds upon the idea of sparse regression, seeking to find a minimal set of nonlinear functions that best describe the system dynamics via an ODE. The SINDy algorithm is relatively easy to implement as it only requires an iterative process of linear regressions. However, one limitation of SINDy, and derivative-based formulations in general, is sensitivity to the error of estimated state time derivatives, where inaccurate derivative estimates can result in the identification of incorrect or suboptimal governing equations. The weak SINDy (wSINDy) method \cite{Messenger2021} addresses this issue by re-formulating the linear regression problem using the so-called weak form.  The weak form uses a set of test functions to replace time derivatives with numerical integrals, thus bypassing the need for the derivative estimates that may be numerically unstable. The wSINDy has been shown to significantly improve model performance in identifying system equations from data by avoiding the amplification of errors resulting from numerical differentiation or noisy measurements \cite{Kaptanoglu2023}. Despite this improvement, wSINDy still heavily relies on a fixed library of functions, usually polynomials, that suffer from the curse of dimensionality. This characteristic fundamentally limits its capacity to model complex, high-dimensional systems, such as the networked systems.

Deep learning-based models can automatically discover low-dimensional latent representations of high-dimensional systems, overcoming the limitations of fixed function libraries and enabling more efficient learning and prediction of complex, nonlinear behaviors \cite{Bengio2013}. Learning dynamics in a latent space offers several key advantages. First, it significantly reduces the dimensionality of the problem, leading to improved computational efficiency, particularly for systems with high-dimensional state spaces or intricate interactions. This reduction allows for faster training and inference, making real-time applications more feasible \cite{Hinton2006}. Second, for high-dimensional dynamical system, by projecting the system dynamics onto a lower-dimensional manifold, latent space representations can capture the essential features and underlying structure of the system, filtering out noise and irrelevant details \cite{Tenenbaum2000}. This focus on core dynamics can lead to models with improved generalizability and predictive accuracy.
There is indeed an extension of wSINDy to a latent dynamics formulation using dimension reduction techniques, such as autoencoders \cite{Tran2024}.  However, in this approach, the dimension reduction and dynamics learning are performed in a sequential manner, and the dynamics still employ a fixed library of functions, which may suffer from lack of expressiveness. 
Furthermore, the latent dynamics extension \cite{Tran2024} did not account for control inputs.

Hence, an alternative continuous-time formulation that is more promising for networked systems is Neural Ordinary Differential Equations (NODEs) \cite{Chen2018neural} and their graph-based variants \cite{Poli2019,Zang2020,Luo2023,Huang2023}.
NODEs are examples of deep learning models that learn continuous-time dynamics in a latent space,
% The NODEs parameterize the time derivative of states using a neural network, allowing for the discovery of complex, nonlinear relationships between the state variables.
and graph-based NODEs extend this idea by incorporating graph structures to capture interactions among the components of a networked system.
The NODE methods belong to another category of continuous-time methods, the ``trajectory-based'' formulation, which fits a model so that the trajectories generated by the model match with those in the dataset.
This formulation results in an ODE-constrained problem that is usually highly nonlinear and non-convex and thus more difficult to solve than the regression problem in the derivative-based formulation.
Furthermore, the gradient computation of NODEs, as needed in the learning process, can be numerically unstable and thus computationally expensive and memory-intensive \cite{Gholami2019anode, Zhuang2020adaptive}. This drawback is particularly pronounced in the case of stiff systems \cite{Kim2021stiff}, which may result in completely incorrect models or even the failure of learning.  Hence the bottleneck of stiffness is highly undesirable for the learning of dynamics for networked systems, which are often stiff due to including multiple timescales.

\paragraph{Contributions}
To bridge the gaps in the literature, we propose a novel and systematic approach for data-driven control-oriented modeling of networked systems. The major contributions of this paper are as follows:
\begin{compactenum}
\item We introduce the weak Latent Dynamics Model (wLDM), a framework for learning generic nonlinear dynamics.  This framework extends the NODE method from the trajectory-based formulation with the weak form concept from wSINDy, i.e., the derivative-based formulation, to eliminate the bottleneck of stiffness in the learning of NODEs and enhance the accuracy of the learned dynamical models.
% a general deep learning framework that leverages a weak form for training. The wLDM offers improved performance and robustness as compared to baseline methods, enabling faster and more stable training even for stiff systems.
\item Building upon the general wLDM framework, we propose the weak Graph Koopman Bilinear Form (wGKBF) model as a control-oriented model for networked systems. The wGKBF model captures the complex dynamics of networked systems while maintaining a relatively simple bilinear structure, leveraging the expressiveness of deep learning for  sparse and heterogeneous network dynamics.
\item We demonstrate the effectiveness of the proposed wLDM method, including its wGKBF variant, through three example systems of increasing complexity: a controlled double pendulum, a stiff Brusselator dynamics model, and an electrified aircraft energy system. The results highlight the capability of the proposed framework to produce accurate and computationally efficient control-oriented models for complex networked systems.
\end{compactenum}

The rest of the paper is organized as follows.  Section \ref{sec:prb} states the general problem of learning dynamics from data and discusses the existing algorithms and their limitations.  Section \ref{sec:ldm} presents the proposed wLDM framework for solving the general problem.  Section \ref{sec:kbf} adapts the wLDM framework to develop wGKBF for the control-oriented modeling of networked systems.  Section \ref{sec:res} demonstrates the efficacy of wLDM and wGKBF models via three numerical examples.  Lastly, the conclusions are provided in Section \ref{sec:con}.

% -----------------------------------------------
\section{Problem Statement and Preliminaries}\label{sec:prb}
% -----------------------------------------------
In this section, we formally state the problem of inferring the parameters of a nonlinear dynamical system from data. Then, we briefly discuss existing algorithms for solving such problems, namely the direct and indirect methods, and their limitations, which serves as the basis for the proposed algorithms.

\subsection{Problem Statement}
Consider a nonlinear dynamical system
\begin{subequations}\label{eqn:system}
\begin{align}\label{eqn:system:a}
\dot{\vx} &= \vf(\vx,\vu;\vtQ) \\
\vy &= \vg(\vx,\vu;\vtQ)   \\\label{eqn:system:c}
\vx(t_1) &= \vx_0(\vtQ)
\end{align}
\end{subequations}
where $\vx\in\bR^{n_x}$ is the state, $\vu\in\bR^{n_u}$ the control, $\vy\in\bR^{n_y}$ the observation, and $\vtQ\in\bR^{n_\theta}$ the model parameters. Functions $\vf$ and $\vg$ can depend on time $t$, however we omit this for simplicity of exposition. The initial condition may be partially or fully unknown, and Eq. \eqref{eqn:system:c} provides an estimate. The goal is to infer parameters $\vtQ$ of the potentially nonlinear models $\vf$ and $\vg$ from a dataset $\cD=\{(\vU^{(i)}, \hvY^{(i)})\}_{i=1}^{n_T}$, where $n_T$ is the total number of trajectories. The $i$th trajectory consists of a sequence of control inputs $\vU^{(i)}=[\vu_1^{(i)},\vu_2^{(i)},\dots,\vu_{L^{(i)}}^{(i)}]\in\bR^{n_u\times L^{(i)}}$ at time stations $\{t_1,t_2,\cdots,t_L\}$ and corresponding measurements $\hvY^{(i)}=[\hvy_1^{(i)},\hvy_2^{(i)},\dots,\hvy_{L^{(i)}}^{(i)}]\in\bR^{n_y\times L^{(i)}}$, where $L^{(i)}$ is the length of the $i$th trajectory.
% , and $\hvy_l^{(i)}=\vy_l^{(i)}+\vth_l^{(i)}$, with $\vth_l^{(i)}$ being the measurement noise at time step $l$.
Without loss of generality, we formulate the optimization problem for learning the dynamics using a single trajectory:
\begin{subequations}\label{eqn:opt}
\begin{align}
\min_\vtQ &\ J(\vtQ;\cD) = \int_{t_1}^{t_L} \norm{\vg(\vx,\vu;\vtQ)-\hvy}^2 dt \\\label{eqn:opt_ode}
\text{s.t.} &\ \dot{\vx} = \vf(\vx,\vu;\vtQ), \quad \vx(t_1) = \vx_0(\vtQ)
\end{align}
\end{subequations}
This is the trajectory-based formulation as categorized in the introduction.

\subsection{Available Algorithms}
To solve Eq. \eqref{eqn:opt}, the continuous-time optimization problem is transformed into a discrete-time approximation through discretization methods, converting the infinite-dimensional problem into a finite-dimensional one.
% When learning nonlinear dynamics from data, the continuous-time optimization problem in Eq. \eqref{eqn:opt} needs to be transformed into a discrete-time approximation to enable numerical solutions. This transformation is achieved through discretization methods, which convert the infinite-dimensional problem into a finite-dimensional one.
There are two main categories of discretization methods \cite{Ross2002}: Direct methods, also known as ``discretize-then-optimize'', and indirect methods, or ``optimize-then-discretize''. These two methods are briefly discussed and compared next.
% The main difference between these two approaches lies in the order of discretization and optimization.

\subsubsection{Direct Methods}
In the direct approach, the differential constraints are discretized, and the continuous-time dynamics and the optimization problem are transformed into a discrete form. This discretization process typically involves approximating the system dynamics at a series of discrete time points using numerical integration methods. For illustrative purposes, applying a simple first-order discretization to Eq. \eqref{eqn:opt} yields a discretized form,
\begin{subequations}
\begin{align}
\min_\vtQ &\ J_d(\vtQ;\cD) = \sum_{i=1}^{L} \norm{\vg(\vx_i,\vu_i;\vtQ)-\hvy_i}^2 \Delta t \\\label{eqn:opt_discrete}
\text{s.t.} &\ \frac{\vx_i-\vx_{i-1}}{\Delta t} = \vf(\vx_i,\vu_i;\vtQ),\ i=2,\dots,L, \quad \vx(t_1)=\vx_0(\vtQ)
\end{align}
\end{subequations}
where $\Delta t$ is the time step between consecutive measurements, and $\vx_i$ and $\vu_i$ represent the state and control input at the $i$-th time step, respectively. By discretizing the continuous-time problem, the direct approach transforms the dynamical system into a set of algebraic equations, which can be solved using standard numerical optimization techniques. The optimization process seeks the parameters $\vtQ$ that minimize the objective function $J_d(\vtQ;\cD)$ while satisfying the discretized dynamics constraints. 

% The direct approach offers several advantages, including the ability to leverage robust numerical optimization algorithms 
% a natural regularization effect due to the discretized dynamics constraints \todo{Add citation}, 
% and the straightforward incorporation of additional constraints.
In practice, more sophisticated discretization schemes, such as pseudospectral methods \cite{Patterson2014}, could be used to reduce the discretization error. However, due to the inevitable discretization errors in Eq. \eqref{eqn:opt_discrete}, the direct method can suffer from the accumulation of approximation errors over time, which may lead to suboptimal solutions. The accuracy and computational efficiency of the optimization process depends heavily on the choice of the discretization scheme and the time step size.

\subsubsection{Indirect Methods}\label{sec:indirect}
Conversely, the indirect approach starts with the necessary optimality conditions of the continuous-time optimization problem 
before any discretization. The optimality conditions are typically obtained by applying Pontryagin's Maximum Principle (PMP) \cite{Pontryagin2018mathematical}, which introduces the adjoint state $\vtl(t)\in\bR^{n_x}$ and the Hamiltonian $H(\vx, \vu, \vtl, t) = \vtl^T \vf(\vx, \vu; \vtQ) + l(\vx,\vu;\vtQ)$, with $l(\vx,\vu;\vtQ) = \norm{\vg(\vx,\vu;\vtQ)-\hvy}^2$.
PMP states that for an optimal solution, the following conditions must hold:
% \begin{subequations}
% \begin{align}
% \dot{\vx} &= \frac{\partial H}{\partial \vtl} = \vf(\vx, \vu; \vtQ) \\\label{eqn:adj}
% \dot{\vtl} &= -\frac{\partial H}{\partial \vx} = -\vtl^T \frac{\partial \vf}{\partial \vx}(\vx, \vu; \vtQ) \\
% \frac{\partial H}{\partial \vu} &= \vtl^T \frac{\partial \vf}{\partial \vu}(\vx, \vu; \vtQ) = 0
% \end{align}
% \end{subequations}
\begin{subequations}\label{eqn_tpbvp}
\begin{align}
\dot{\vx} &= \frac{\partial H}{\partial \vtl} = \vf(\vx, \vu; \vtQ), \quad\vx(t_1) = \vx_0\\\label{eqn:adj}
\dot{\vtl} &= -\frac{\partial H}{\partial \vx} = -\vtl^T \frac{\partial \vf}{\partial \vx}(\vx, \vu; \vtQ) - \frac{\partial l}{\partial \vx}(\vx, \vu; \vtQ), \quad\vtl(t_L) =  \frac{\partial J}{\partial \vx(t_L)}\\
0 &= \frac{\partial H}{\partial \vu} = \vtl^T \frac{\partial \vf}{\partial \vu}(\vx, \vu; \vtQ) + \frac{\partial l}{\partial \vu}(\vx, \vu; \vtQ)
\end{align}
\end{subequations}
where Eq. \eqref{eqn:adj} governs the evolution of the adjoint state, and is solved backward in time with the terminal condition of $\frac{\partial J}{\partial\vx(t_L)}$; the value of $l$ at any time may be interpolated from the given discrete measurements.  Equation \eqref{eqn_tpbvp} defines a two point boundary value problem.

There are two popular approaches for solving the boundary value problem: The shooting method and the collocation method \cite{Ross2015}. The shooting method iteratively solves the problem by guessing the initial adjoint state and updating it based on the mismatch between the final state and the desired terminal condition. The collocation method discretizes the continuous-time optimality conditions using numerical schemes and solves the resulting nonlinear algebraic equation. One of the main advantages of the indirect approach is that it inherently handles the continuous nature of the dynamics and adjoint equations, leading to more accurate gradients and solutions as compared to the direct approach.  However, the indirect approach often suffers from numerical difficulties during optimization, such as the need for good initial guesses for the adjoint state and system parameters, and the sensitivity to the choice of numerical schemes for discretizing the optimality conditions.

\subsubsection{Applications in Machine Learning}

In the context of machine learning, when the trajectory optimization problem is chosen to learn dynamics from data, the indirect method is usually selected due to its guarantee optimality guarantees. In this case, the learning algorithm falls under the category of NODEs.  The training of NODEs has been implemented using shooting methods \cite{Chen2018neural}. While NODEs provide a continuous-time modeling framework and convert the constrained optimization problem into an unconstrained one, training them can be computationally expensive and memory-intensive for solving the adjoint ODE.  Particularly for stiff systems, solution of the adjoint ODE may be numerically unstable and can even lead to failure in convergence of model learning \cite{Kim2021stiff}. Moreover, the numerical integration process can introduce approximation errors and lead to numerical instabilities.  These challenges in existing algorithms motivate the exploration of alternative approaches, such as the weak form, as discussed in the following section.

% -----------------------------------------------
\section{Formulation of Weak Latent Dynamics Model}\label{sec:ldm}
% -----------------------------------------------

In this section, we introduce the weak form approach and propose the weak Latent Dynamics Model (wLDM), a general deep learning architecture that learns the dynamics of a system in a latent space. We detail the key components of the wLDM and discuss its advantages over conventional methods.

\subsection{Weak Form for Learning Dynamics}
% The weak form is a discretization method that generalizes the derivative-based formulation and reformulates the optimization problem by converting the time derivatives into integrals over time intervals using test functions.
% In the context of the NODE methods, i.e., trajectory-based formulations, the concept of the weak form is adopted to bypass the integral constraints. 

The original formulation of weak form for learning dynamics in Ref. \cite{Messenger2021} is developed for derivative-based formulation assuming that the states are known; this results in a linear regression problem, possibly with regularization terms.  We extend the approach to the NODE methods, i.e., trajectory-based formulation, where the states are unknown. This results in a nonlinear regression problem, where the mapping between the physical and latent states as well as the latent dynamics are learned together in an iterative manner.
Essentially, the NODE methods are reformulated using the weak form to provide a computationally {efficient and numerically stable} framework for identification of the nonlinear dynamics.

\subsubsection{Representation of Dynamics by Weak Form}\label{sec:weak}
Consider a set of $C^1$-continuous test functions $\{\phi_i(t)\}_{i=1}^{n_\phi}$.  A test function $\phi(t)$ is defined on a time interval $[a,b]$, such that $\phi(t)=0$ for $t\notin[a,b]$ and $\phi$ satisfies the boundary conditions:
\begin{equation}\label{eqn:tst_bc}
\phi(a)=\phi(b)=0
\end{equation}

The ODE for the dynamics in Eq. \eqref{eqn:system} can be converted to the weak form by multiplying both sides of the equation by a test function $\phi(t)$ and integrating over the time interval $[a,b]$:
\begin{equation}\label{eqn:wf_int}
\int_a^b \phi(t) \dot{\vx}(t) dt = \int_a^b \phi(t) \vf(\vx,\vu;\vtQ) dt
\end{equation}
Using integration by parts and the boundary conditions in Eq. \eqref{eqn:tst_bc}, we can eliminate the time derivative of $\vx$ on the left-hand side of Eq. \eqref{eqn:wf_int}:
\begin{equation}\label{eqn:wf_ibp}
\int_a^b \phi(t) \dot{\vx}(t) dt = \phi(t)\vx(t)\Big|_a^b - \int_a^b \dot{\phi}_i(t) \vx(t) dt = - \int_a^b \dot{\phi}_i(t) \vx(t) dt
\end{equation}
Substituting Eq. \eqref{eqn:wf_ibp} into Eq. \eqref{eqn:wf_int}, we obtain the weak form of the dynamics:
\begin{equation}\label{eqn:wf}
-\int_a^b \dot\phi(t) \vx(t) dt = \int_a^b \phi(t) \vf(\vx,\vu;\vtQ) dt
\end{equation}
The key difference of Eq. \eqref{eqn:wf} when compared to the original dynamics in Eq. \eqref{eqn:system} is that $\dot{\vx}$ no longer appears. This would have to be estimated from the data of $\vx$ and the estimation error becomes part of the model error that lead to incorrect learning of dynamics.
In the weak form of  Eq. \eqref{eqn:wf}, the time derivative is transferred to the test function $\phi$ that is chosen by the user and the exact expression of $\dot{\phi}$ is known. Hence, the use of a test function bypasses the need to estimate $\dot{\vx}$ and eliminates the model error due to derivative estimation.

The weak form in Eq. \eqref{eqn:wf} allows us to reformulate the optimization problem in Eq. \eqref{eqn:opt} as:
\begin{subequations}\label{eqn:opt_int}
\begin{align}\label{eqn:opt_obj}
\min_{\vtQ} &\ J(\vtQ;\cD) = \int_{t_1}^{t_L} \norm{\vg(\vx,\vu;\vtQ)-\hvy}^2 dt \\\label{eqn:opt_ic}
\text{s.t.} &\ - \int_{a_i}^{b_i} \dot{\phi}_i(t) \vx(t) dt = \int_{a_i}^{b_i} \phi_i(t) \vf(\vx,\vu;\vtQ) dt, \quad i=1,\dots,n_{\phi}
\end{align}
\end{subequations}

As the number of test functions $n_\phi$ approaches infinity and the union of intervals $[a_i,b_i]$ cover $[t_1,t_L]$, the weak form constraints in Eq. \eqref{eqn:opt_ic} converge to the strong form of the dynamics, resulting in exact satisfaction of the system dynamics. However, when working with a finite number of test functions, Eq. \eqref{eqn:opt_ic} only ``weakly'' satisfies the true dynamics, hence the name ``weak form''. The accuracy of the approximation depends on a number of factors that are discussed later in Sec. \ref{sec:choice}.

\subsubsection{Discretization and Integration Matrices}
Next, we discretize the integrals in Eq. \eqref{eqn:opt_int} based on the dataset $\cD$, and convert the optimization to a conventional equality-constrained problem. For any integrable function $\eta(t)$, given data pairs $\{(t_k,\eta(t_k))\}_{k=1}^N$, $a\leq t_1<t_2<\cdots<t_N\leq b$, we can perform numerical integration as follows:
\begin{equation}
    \int_a^b \eta(t) dt \approx \sum_{i=1}^N v_k\eta(t_k) \equiv \vth^T\mathbf{v}
\end{equation}
where $v_k$ are weights at time stations $t_k$, and
\begin{gather}
    \vth=[\eta(t_1),\eta(t_2),\cdots,\eta(t_N)]^T,\quad \mathbf{v}=[v_1,v_2,\cdots,v_N]^T
\end{gather}
Here the number of steps $N$ in the interval of the integral will be referred to as the ``time horizon'' for conciseness. % This is not to be confused with time horizon in MPC methods.

% Using numerical integration, the objective function Eq. \eqref{eqn:opt_obj} is approximated as
% \begin{align}\nonumber
%     J(\vtQ;\cD) &= \int_{t_1}^{t_L} \norm{\vg(\vx,\vu;\vtQ)-\hvy}^2 dt \\\nonumber
%     &\approx \sum_{k=1}^K v_k\norm{\vg_k-\hvy_k}^2 \\\label{eqn:wf_obj}
%     &\equiv \norm{Z-\tvY}_v^2 
% \end{align}
% where Eq. \eqref{eqn:vnorm} is a weighted Frobenius norm, denoted as
% \begin{equation}
%     \norm{Z-\tvY}_v^2 = \tr((\vY-\tvY)\mathbf{v}(\vY-\tvY)^T)
% \end{equation}
% where $\vg_k=\vg(\vx_k,\vu_k;\vtQ)$, $\vY=[\vg_1,\vg_2,\cdots,\vg_K]$, $\tvY=[\tvy_1,\tvy_2,\cdots,\tvy_K]$, $\vV=\diag(\mathbf{v})$, and $\tr()$ denotes matrix trace.

The objective function in Eq. \eqref{eqn:opt_obj} can be approximated using numerical integration and the weighted Frobenius norm $\norm{\cdot}_v$ as follows:
\begin{align}
J(\vtQ;\cD) &= \int_{t_1}^{t_L} \norm{\vg(\vx,\vu;\vtQ)-\hvy}^2 dt \\\nonumber\label{eqn:wf_obj}
&\approx \sum_{k=1}^N v_k\norm{\vy_k-\hvy_k}^2 \equiv \norm{\vY-\hvY}_v^2 
\end{align}
where $\vy_k=\vg(\vx_k,\vu_k;\vtQ)$.
The weighted Frobenius norm is defined as:
\begin{equation}
\norm{\vY-\hvY}_v^2 = \tr((\vY-\hvY)\mathbf{V}(\vY-\hvY)^T)
\end{equation}
where $\vY=[\vy_1,\vy_2,\dots,\vy_N]$, $\hvY=[\hvy_1,\hvy_2,\dots,\hvy_N]$, $\mathbf{V}=\diag(v)$, and $\tr(\cdot)$ denotes the matrix trace.

Similarly, for the constraints in Eq. \eqref{eqn:opt_ic}, the left-hand side of the $i$th constraint is approximated as
\begin{equation}
    - \int_a^b \dot{\phi}_i\vx dt \approx -\sum_{k=1}^N v_k\dot{\phi}_i(t_k)\vx_k \equiv \vX\vd_i
\end{equation}
where $\vX=[\vx_1,\vx_2,\cdots,\vx_N]$,
$
    \vd_i = [v_1\dot{\phi}_i(t_1),v_2\dot{\phi}_i(t_2),\cdots,v_N\dot{\phi}_i(t_N)]^T
$,
and the right hand side is
\begin{equation}
    \int_a^b \phi_i\vf(\vx,\vu;\vtQ) dt \approx -\sum_{k=1}^K v_k\phi_i(t_k)\vf_k \equiv \vF\vc_i
\end{equation}
where $\vf_k=\vf(\vx_k,\vu_k;\vtQ)$, $\vF=[\vf_1,\vf_2,\cdots,\vf_N]$ and
$
    \vc_i = [v_1{\phi}_i(t_1),v_2{\phi}_i(t_2),\cdots,v_N{\phi}_i(t_N)]^T
$.

The total of $n_\phi$ integral constraints can be written compactly as
\begin{equation}\label{eqn:wf_cons}
    \vX\vD = \vF(\vX,\vU;\vtQ)\vC
\end{equation}
where $\vD=[\vd_1,\vd_2,\cdots,\vd_{n_\phi}]$, $\vC=[\vc_1,\vc_2,\cdots,\vc_{n_\phi}]$, and  the dependence of $\vF$ on states, inputs, and parameters is explicitly written. 

The discretized optimization problem with algebraic equality constraints takes the form:
\begin{subequations}\label{eqn:opt_wf}
\begin{align}\label{eqn:opt_wf_obj}
\min_\vtQ &\ \norm{\vY-\hvY}_v^2 \\
\text{s.t.} &\ \vX\vD = \vF(\vX,\vU;\vtQ)\vC\label{eqn:opt_wf_con}
\end{align}
\end{subequations}
To solve the constrained optimization problem in Equation \eqref{eqn:opt_wf}, 
% one approach is to use the Lagrangian method, which introduces Lagrange multipliers to incorporate the equality constraints into the objective function. However, 
we introduce a penalty for constraint violations, resulting in the following loss function:
\begin{equation}\label{eqn:loss}
\cL(\vtQ) = \norm{\vY-\hvY}_v^2 + \gamma \norm{\vX\vD - \vF(\vX, \vU; \vtQ)\vC}^2,    
\end{equation}
where $\gamma$ is the penalty parameter. More advanced methods such as the interior point method may be used for more accurate enforcement of the constraints.

\subsubsection{Choice of Test Functions}\label{sec:choice}
The choice of test functions $\phi_i(t)$ is crucial to the performance of the weak form. In Ref. \cite{Messenger2021} the test functions are from the space of piece-wise polynomials that are non-negative, unimodal, and compactly supported in $[0,T]$ with a certain number of continuous derivatives depending on the polynomial degrees.

In this work, we generalize the approach and propose using test functions based on Jacobi polynomials $P_n^{(\alpha,\beta)}(\tau)$, which are orthogonal on the interval $\tau\in[-1,1]$ with respect to the weight function $\rho^{(\alpha,\beta)}(\tau)=(1-\tau)^\alpha(1+\tau)^\beta$. The orthogonality property of Jacobi polynomials is given by:
\begin{equation}
    \int_{-1}^1 P_n^{(\alpha,\beta)}(\tau)P_m^{(\alpha,\beta)}(\tau)\rho^{(\alpha,\beta)}(\tau) d\tau = K_n^{(\alpha,\beta)} \delta_{nm}
\end{equation}
where $K_n^{(\alpha,\beta)}$ is a normalization constant and $\delta_{nm}$ is the Dirac delta.
We define the test functions $\phi_i(t)$ using Jacobi polynomials and the weight function $\rho^{(1,1)}(\tau)$ as follows:
\begin{equation}
\phi_i(t) = P_i^{(1,1)}(\tau)\rho^{(1,1)}(\tau), \quad \tau = \frac{2t-(b+a)}{b-a}
\end{equation}
The variable transformation $\tau$ maps the time interval $[a,b]$ to the standard interval $[-1,1]$. The weight function $\rho^{(1,1)}(\tau)=(1-\tau^2)$ ensures that the boundary conditions in Eq. \eqref{eqn:tst_bc} are satisfied, as $\phi_i(a_i)=\phi_i(b_i)=0$. The order of the Jacobi polynomial also becomes a tunable hyperparameter of the weak form.

The use of Jacobi polynomials as test functions offers several potential advantages over the original choice in Ref. \cite{Messenger2021}. First, the orthogonality property of Jacobi polynomials can lead to better convergence in representing the system trajectories as compared to non-orthogonal basis functions. Second, Jacobi polynomials are smooth functions, which can help capture the underlying dynamics more accurately. Third, the weight function $\rho^{(1,1)}(\tau)$ automatically enforces the boundary conditions in Eq. \eqref{eqn:tst_bc}, simplifying the implementation of the weak form. Furthermore, the choice of test functions based on Jacobi polynomials allows for the accurate approximation of the integral constraints using quadrature rules. 
% As shown in Ref. \cite{Messenger2021}, the trapezoidal rule can discretize the weak derivative relation Eq. \eqref{eqn:wf} to a high order of accuracy \todo{show actual order} when the test functions have roots of sufficient multiplicity at the endpoints of their support. 
As shown in Ref. \cite{Messenger2021}, the trapezoidal rule can discretize the weak derivative relation in Eq. \eqref{eqn:wf} to order $p+1$ accuracy when the test function $\phi$ has roots $\phi(a)=\phi(b)=0$ of multiplicity $p$ at the endpoints of its support. This leads to more accurate representation of the dynamics and potentially a better model.

The choice of the time horizon $N$, polynomial order (PO), and integration order (IO) in the weak form presents a delicate balance between approximation accuracy and computational efficiency. The time horizon $N$ determines the length of the trajectory segment considered by each constraint; an intuition is that a longer segment captures more dynamics within the time interval and results in a more accurate characterization of dynamics by the constraint.
% and an increase in $N$ reduces the number of integral constraints.
The PO affects the expressiveness of the test functions $\phi_i(t)$ used in the weak form, with higher-order polynomials providing more flexibility, and hence potentially more accuracy, in representing complex dynamics. The IO, which corresponds to the accuracy of the numerical integration scheme, can be improved using higher-order methods, such as Simpson's rule or higher-order Newton-Cotes formulas. Balancing these hyperparameters is essential for accurately and efficiently learning dynamical systems using the weak form.  In particular, while larger values of $N$ and PO may result in a more accurate dynamic model representation, but the increased $N$ and PO would increase the dimensions of weight matrices $\vC$ and $\vD$ in the integral constraints, Eq. \eqref{eqn:opt_wf_con}, and thus increase the computational cost in the evaluation of these constraints during the model training.
% The optimal choice depends on the system's dimensionality, nonlinearity, and complexity.
In general, cross-validation or other model selection techniques can be used to tune these hyperparameters for the best trade-off between accuracy and computational efficiency. The effects of $N$, PO, and IO on the weak form's performance will be investigated through numerical examples in Sec. \ref{sec:dp}, providing empirical evidence of their impact on model accuracy and guidance in the choice of these parameters.

% Intuitively, a larger $N$ allows the model to capture more information about the system dynamics by enforcing the ODE constraints at more points along the trajectory.
% The interplay between $N$, PO, and IO is complex, as a higher PO or IO can potentially reduce the required $N$ to achieve a desired level of accuracy, while a lower PO or IO may necessitate a larger $N$ to maintain the same approximation quality.

\subsection{Weak Latent Dynamics Model}\label{sec:wldm}
Building upon the weak form approach introduced in the previous section, we propose the weak Latent Dynamics Model (wLDM), a deep learning-based general architecture that learns the dynamics of a system in a latent space. The wLDM leverages the advantages of the weak form to provide an efficient and stable framework for learning complex, nonlinear dynamics from data.
The wLDM consists of three main components: An encoder, a processor, and a decoder, as illustrated in Fig. \ref{wLDM}. These components work together to map the observed data into a latent space, learn the dynamics in this latent space, and reconstruct the original system states from the latent representations.
The components of the wLDM are detailed as follows:

\begin{compactenum}
    \item Encoder: The encoder $\vf_E$ maps the observations $\vy$ and control inputs $\vu$ into a $S$-dimensional latent space representation $\vw\in\bR^S$ as:
    \begin{equation}\label{eqn:encoder}
        \vw = \vf_E(\vy, \vu;\vtQ_E) 
    \end{equation}
    where $\vtQ_E$ denotes the learnable parameters of the encoder.
    This latent representation is intended to capture the essential features and dynamics of the system in a compact and computationally amenable form.
    The choice of the encoder architecture depends on the nature of the data and the problem at hand. For instance, fully connected neural networks (FCNNs) can be used for simple, structured data, while graph neural networks (GNNs) may be more suitable for data with complex, graph-like structures. The encoder consists of a neural network of $N_E$ layers.
    
    \item Processor: The processor $\vf_P$ is a neural network of $N_P$ layers that learns the dynamics in the latent space and produces the time derivative of the latent states, $\dot{\vw}\in\bR^S$ as
    \begin{equation}
        \dot{\vw} = \vf_P(\vw, \vu; \vtQ_P)
    \end{equation}
    where $\vtQ_P$ represents the learnable parameters of the processor. 
    The processor plays a crucial role in the wLDM, as it learns the underlying dynamics of the system in the latent space. By operating in a possibly lower-dimensional space, the processor can learn more efficient and stable representations of the dynamics as compared to learning directly in the original state space.
    
    \item Decoder: The decoder $\vf_D$ is a neural network of $N_D$ layers, which reconstructs the observations from the latent representation $\vw$,
    \begin{equation}
        \vy = \vf_D(\vw; \vtQ_D)
    \end{equation}
    where $\vtQ_D$ represents the learnable parameters of the decoder.
% The decoder ensures that the learned latent dynamics can be mapped back to the original state space, enabling the wLDM to generate predictions and reconstruct the observed data.
\end{compactenum}

% \todo{Yin - update the symbols to be lower case, since we are not using TDE.  Also, think about Dr. Pangborn's comment.}

\insertfig{wLDM}{0.6}{Structural diagram of the wLDM architecture.}
Each component of the wLDM is typically composed of multiple layers of neural networks, with nonlinear activation functions applied between layers. A common choice for the activation function is the Parametric Rectified Linear Unit (PReLU), which allows for learning the optimal slope of the negative part of the activation function during training. The choice of latent space dimensionality $S$ is a crucial hyperparameter that balances the trade-off between model complexity and information loss. A smaller $S$ leads to a more compact representation but may result in a loss of information, while a larger $S$ can capture more complex dynamics but may increase computational cost and the risk of overfitting. In practice, the optimal value of $S$ can be determined through cross-validation or based on prior knowledge of the system's intrinsic dimensionality. The wLDM is trained by minimizing the loss function in Eq. \eqref{eqn:loss} using standard optimization algorithms, such as Adam, with an exponential decay scheduling of the learning rate to ensure stable convergence.
Another set of critical hyperparameters in the wLDM are the choice of intervals $[a_i,b_i]$, i.e., their lengths and distribution.  Empirically, the intervals are chosen to share the same length, characterized by the number of time steps within the interval $N$; the intervals are distributed evenly and closely so that neighboring intervals overlap by at least $N/4$.

The trained model can then make predictions of system dynamics given initial conditions and control inputs, as summarized in Algorithm \ref{algo:pred}. {Note that the algorithm leverages the decoding of the predicted observations at each time step. By encoding the predicted observations into the latent space using the encoder $\vf_E$ and then decoding the updated latent states back to the original space using the decoder $\vf_D$, the algorithm ensures that the predicted trajectories remain within the valid and meaningful region of the state space captured by the encoder. This decoder step helps to maintain the stability of the predictions, reduces the influence of noise, and improves the generalization capability of the model. The benefits of this approach have been empirically observed, with the decoder-based predictions consistently leading to more accurate results as compared to evolving the dynamics solely in the latent space without intermediate encoding and decoding steps.}

\begin{algorithm}
\caption{Predict System Dynamics Using Trained Model}
\begin{algorithmic}[1]\label{algo:pred}
\REQUIRE Trained wMLP with learned parameters $\vtQ$
\REQUIRE Initial observation $\hvy_0$, control inputs $\vu(t)$, prediction horizon $T$
\STATE Encode initial observation: $\vw_0 = \vf_E(\hvy_0)$
\STATE Initialize predicted trajectory: $\vy(0) = \hvy_0$
\STATE Initialize time step: $t = 0$
\WHILE{$t < T$}
    \STATE Encode current observation: $\vw_t=\vf_E(\vy_t, \vu_t)$
    \STATE Compute latent state derivative: $\dot{\vw}_t = \vf_P(\vw_t, \vu_t; \vtQ_P)$
    \STATE Update latent state using ODE solver: $\vw_{t+\Delta t} = \text{ODESolver}(\vw_t, \dot{\vw}_t, \Delta t)$
    \STATE Decode latent state: $\vy_{t+\Delta t} = \vf_D(\vw_{t+\Delta t})$
    % \STATE Append predicted observation to trajectory: $\vy_{t+\Delta t} \rightarrow \vy_t$
    \STATE Update time step: $t = t + \Delta t$
\ENDWHILE
\RETURN Predicted trajectory $\vy(t)$
\end{algorithmic}
\end{algorithm}

The key advantage offered by wLDM over conventional methods such as NODEs is that the former does not rely on the solution of adjoint ODE for gradient computation during the training process, which can be computationally expensive and memory-intensive, especially for long sequences, high-dimensional systems, or stiff systems. Using the weak form, the wLDM can be trained with significantly less computational cost, which will be demonstrated in the results section. In addition, the wLDM architecture, as illustrated in Fig. \ref{wLDM}, is flexible in accommodating different types of neural network layers for the encoder, processor, and decoder, depending on the specific problem and the structure of the data. For instance, FCNNs can be used for simple, structured data, while GNNs may be more suitable for data with complex, graph-like structures.

\section{Control-Oriented Modeling of Networked Systems}\label{sec:kbf}
% -----------------------------------------------
In this section, we introduce the weak Graph Koopman Bilinear Form (wGKBF), which is a specific adaptation of the wLDM framework tailored to networked systems.
We start by stating the model problem and the corresponding KBF model for networked systems, subsequently present GNNs as the key component for incorporating graph topology into the wLDM, and lastly present the full wGKBF model under the wLDM framework.

\subsection{Model Problem}\label{sec:gdyn}

In this study, we consider a networked system whose components are sparsely connected and each may have different numbers of states and inputs.  While this system is physically heterogeneous, a homogeneous graph will be used to characterize the topology, i.e., the nodes and the edges.  The heterogeneous information on the nodes, e.g., the states and inputs, will be homogenized later in the wGKBF model.

First, the notations for graphs are briefly introduced.
Let $\cG=(\cV,\cE)$ be a graph with a set of $n_N$ nodes, $\cV\subset\bN^+$, and a set of edges $\cE\subseteq\cV\times\cV$.  The graph may contain directed edges between nodes $I$ and $J$, i.e., $(I,J)\in\cE$ and $(J,I)\notin\cE$.  By convention, $(I,J)$ means $I$ pointing to $J$.  Note that capitalized indices are used to differentiate from indices for time steps.
For node $I$, define its neighbors as $\cN_{I}=\{K|(K,I)\in\cE\}$.  The number of neighbors, $\abs{\cN_{I}}$, is referred to as the degree of node $I$, $d_I$.
As will be used later, the connectivity of a graph is characterized by the adjacency matrix $\vS\in\bR^{n_N\times n_N}$, with $\vS_{IJ}=1$ if $(I,J)\in\cE$ and 0 otherwise.  Furthermore, the (normalized) Laplacian matrix is defined as $\vL=\vI-\vD^{-1/2}\vS\vD^{-1/2}$, where $\vD$ is the diagonal matrix with $\vD_{II}=d_I$.

Next, the dynamics of a networked system on graph $\cG$ are defined.
Suppose node $I$ has states $\vx_I\in\bR^{n_I}$ and inputs $\vu_I\in\bR^{m_I}$, and the state and input dimensions can be different among the nodes.  The dynamics at node $I$ are governed by the following system of ODEs:
\begin{equation}\label{eqn_nod}
    \dot{\vx}_I = \vf_I(\vx_I,\vu_I) + \sum_{J\in\cN_{I}} \vf_{IJ}(\vx_I,\vx_J,\vu_I)
\end{equation}
where the first term accounts for the self-interaction dynamics, with respect to $\vx_I$ and $\vu_I$, and the summation accounts for pair-wise interactions due to the neighbors.

The full system may be denoted compactly in the generic form Eq. \eqref{eqn:system:a}, where the collections of states and inputs of the networked dynamics are, respectively,
$$
    \vx = [\vx_1,\vx_2,\cdots,\vx_{n_N}]\in\bR^{n_x},\quad \vu = [\vu_1,\vu_2,\cdots,\vu_{m_N}]\in\bR^{n_u}
$$
with $n_x=\sum_{I=1}^{n_N} n_I$ and $n_u=\sum_{I=1}^{n_N} m_I$.

Koopman theory and its extension to control \cite{Proctor2018,Goswami2022} state that there exists a pair of nonlinear mappings: (1) a lifting, or effectively an encoder, $\vw=\vf_E(\vx,\vu)$ that maps physical states and inputs to the Koopman states $\vw$, and (2) an inverse map, or effectively a decoder, $\vx=\vf_D(\vw)$ that recovers physical states from Koopman states.  In the Koopman space, the dynamics evolve in a linear manner, $\dot{\vw}=\vA\vw$, where $\vA$ is a system matrix to be identified from data or the governing equations.  More details on the Koopman theory are provided in Appendix \ref{app:kbf}.

In this study, we consider control-affine dynamics, for which the simpler KBF model \cite{Goswami2022} may be applied,
\begin{equation}
    \dot{\vw} = \vA\vw + \sum_{k=1}^{n_u}\vB_k\vw u_k
\end{equation}
In the general case where the states and inputs are nonlinearly mixed, the Koopman model may be obtained via, e.g., the KIC framework \cite{Proctor2018}; however, such models might not have convenient structures like the KBF that can be leveraged for controller design.

To learn the KBF model, two elements need to be determined: (1) The nonlinear mappings $\vf_E$ and $\vf_D$, and (2) the bilinear system matrices $\vA$ and $\vB_k$.  Intuitively, the Koopman states for a node $I$ should depend on only the physical states of node $I$ and its neighbors, since distant nodes should have negligible impact on the node $I$.  To preserve such sparsity induced by the graph topology, GNNs will be employed to learn the nonlinear mappings, as will be discussed in Sec. \ref{sec:gnn}.  Subsequently, noting that the Koopman and physical states are effectively the states and observations in the general nonlinear dynamics, respectively, the graph KBF model can be directly learned via the wLDM framework; this will be discussed in Sec. \ref{sec:wgkbf}.

\subsection{Graph Neural Networks}\label{sec:gnn}

The core model structure in GNNs that addresses the graph topology is the message passing (MP) mechanism, which may consist of multiple consecutive MP steps \cite{Hamilton2020}. At the $j$th MP step on a graph $\cG$, a vector of features $\vh_I^{(j)}\in\bR^{n_D^{(j)}}$ at a node $I$ is updated via the following rules,
\begin{subequations}
\begin{align}\label{eqn:ebd}
  &\text{Aggregate:}\quad \vm_I^{(j)} = \vth_{ag}\left(\{\vh_J^{(j)}\ |\ J\in\cN_I\}\right), \\
  &\text{Update:}\quad \vh_I^{(j+1)} = \vth_{up}\left(\vh_I^{(j)},\vm_I^{(j)}\right),
\end{align}
\end{subequations}
where $\vth_{ag}$ and $\vth_{up}$ are nonlinear functions, e.g., neural networks, to be learned from data, and $\vm_I^{(j)}$ denotes the information aggregated from the neighborhood $\cN_I$.  Hence the new vector of features at node $I$, $\vh_I^{(j+1)}\in\bR^{n_D^{(j+1)}}$, is a function of the features from its neighbors.  In essence, one MP step facilitates the exchange of information between directly connected nodes. By stacking multiple MP steps, information can be propagated across wider neighborhoods, allowing each node's feature vector to be influenced by its more distant neighbors.

In this study, the MP mechanism is implemented using Graph Convolution Layers (GCLs) \cite{Defferrard2017}.
GCLs generalize convolution from multi-dimensional data arrays to data on graphs and serve as an effective filter that extracts localized features from graph data. From the point of view of graph Fourier Transforms, GCLs perform filtering on the spectrum of the Laplacian matrix $\vL$ of graph $\cG$.
An efficient implementation of the GCL is the ChebConv network \cite{Defferrard2017}, which performs the MP aggregation and updating over all nodes simultaneously. Collecting the vectors of features at all nodes as $\vH^{(j)}=[\vh_1^{(j)}, \vh_2^{(j)}, \cdots, \vh_{n_N}^{(j)}]^T\in\bR^{n_N\times n_D^{(j)}}$, we define the $K$th order ChebConv-based graph convolution as
\begin{equation}\label{eqn:cheb1}
    \vH^{(j+1)} = \vts\left( \sum_{k=0}^{K}T_k(\tvL)\vH^{(j)}\vtQ_k^{(j)} \right),
\end{equation}
where $\vts$ is a nonlinear element-wise activation function, $\vtQ_k^{(j)}\in\bR^{n_D^{(j)}\times n_D^{(j+1)}}$ are learnable parameters, and in the summation $T_k(\tvL)$ is the $k$th order Chebyshev polynomial of the scaled Laplacian $\tvL$.
The scaled Laplacian is defined as $\tvL = (2/\lambda_{\max})\vL-\vI$ with $\lambda_{\max}$ the largest eigenvalue of $\vL$, so that the range of eigenvalues of $\tvL$ is $[-1,1]$ and matches with the domain of Chebyshev polynomials.  The matrix $T_k(\tvL)$ contains up to $K$th power of $\tvL$ and hence one $K$th order ChebConv in Eq. \eqref{eqn:cheb1} performs $K$ MP steps.

% In the context of the aircraft energy system model used to demonstrate the proposed approach in Section \ref{sec:aircraft}, GNNs serve as powerful graph encoders, efficiently processing the graph topology information and learning compact representations of the system's dynamics. By leveraging the GCL operator, the GNN can aggregate information from neighboring nodes and generate embeddings that capture the intricate relationships and dependencies within the system. These learned embeddings can then be utilized in the KBF to model the evolution of the system's states over time.

\subsection{Weak Graph Koopman Bilinear Form}\label{sec:wgkbf}
Building upon the principles of Koopman theory (Sec. \ref{sec:gdyn}), and the GNN architecture (Sec. \ref{sec:gnn}), we now introduce the weak Graph Koopman Bilinear Form (wGKBF) model, which incorporates graph representations and KBF to capture the dynamics of control-affine networked systems. The wGKBF model extends the wLDM framework by employing GCLs in the encoder and decoder, while the processor takes the form of a KBF.  The details are explained below,

\begin{compactenum}
  \item Encoder: The encoder embeds the physical states at each node into a $\tilde{S}$-dimensional vector of Koopman states so that the entirety of the physical states are embedded into a Koopman space of $S=n_N\times \tilde{S}$ dimensions.  The encoder starts with a simple FCNN step to map the physical states at each node to a feature vector of dimension $\tilde{S}$.  For node $I$,
  \begin{equation}
      \vh_I^{(0)}=\vf_{E,I}(\vx_I,\vtQ_{E,I})
  \end{equation}
  where $\vtQ_{E,I}$ denotes learnable parameters.
  Depending on the applications, the FCNNs at different nodes may be different due to the heterogeneity in state dimensions; in this case the FCNN step is to ensure the uniformity of dimensions in the GCL layers.  Subsequently, the initial collection of features is formed as
  \begin{equation}
      \vH^{(0)} = [\vh_1,\vh_2,\cdots,\vh_{n_N}]^T \in\bR^{n_N\times \tilde{S}}
  \end{equation}
  and passed through $N_E$ GCL layers to produce $\vH^{(N_E)}=[\vh_1^{(N_E)},\vh_2^{(N_E)},\cdots,\vh_{n_N}^{(N_E)}]^T$, which is assembled into the vector of Koopman states,
  \begin{equation}
      \vw = [\vh_1^{(N_E)T},\vh_2^{(N_E)T},\cdots,\vh_{n_N}^{(N_E)T}]^T \in \bR^{S}
  \end{equation}
  Combining the FCNN and GCL steps, the encoder is denoted
  \begin{equation}
  \vw = \vf_E(\vx,\cG; \vtQ_E),
  \end{equation}
  where $\vtQ_E$ denotes all the learnable parameters in the encoder.
  \item Processor: The processor assumes KBF dynamics from Eq. \eqref{eqn:kbf} in the latent space, where the KBF matrices $\vA\in\bR^{S\times S}$ and $\vB_i\in\bR^{S\times S}$ are trainable together with other network parameters.
  \item Decoder: The decoder recovers the physical states at each node from the Koopman space, which is an inversion of the encoder steps.  First, given a new Koopman state vector $\vw$, it is reorganized into a collection of features at all nodes $\vH^{(0)}\in\bR^{n_N\times \tilde{S}}$, which is subsequently passed through $N_D$ GCL layers, arriving at $\vH^{(N_D)}=[\vh_1^{(N_D)},\vh_2^{(N_D)},\cdots,\vh_{n_N}^{(N_D)}]^T$.  Lastly, a simple FCNN is applied to map feature vector back to the physical states at each node.  For node $I$,
  \begin{equation}
      \vx_I=\vf_{D,I}(\vh_I^{(N_D)},\vtQ_{D,I})
  \end{equation}
  Again, these FCNNs at different nodes may be different due to the heterogeneity in state dimensions.
  Combining the GCL and FCNN steps, the decoder is denoted
  \begin{equation}
  \vx = \vf_D(\vw,\cG; \vtQ_D),
  \end{equation}
  where $\vtQ_D$ denotes all the learnable parameters in the decoder.
\end{compactenum}
When compared to the general wLDM framework, the encoder in wGKBF does not treat the inputs. Instead, the inputs are  provided directly to the processor following the KBF formulation.
Finally, when the physical states at each node share the same dimension and physical meaning, the FCNN steps in the encoder and decoder may be skipped.

% Given the latent representation $\vW$ and the control inputs $\vU$, the processor $\vf_P$ in Eq. \eqref{eqn:processor} is adapted as:
% \begin{equation}
% \dot{\vw}_k = \vA\vw_k + \sum_{i=1}^m \vB_i\vw_k u_i,
% \end{equation}

The incorporation of graph-based representations and the KBF in the wGKBF model allows for the effective modeling of networked systems, capturing the intricate interactions and dependencies among the system components. The GCL layers enable the model to learn localized features from the graph structure while the KBF processor provides a global bilinearization of the dynamics in the latent space. By leveraging the expressive power of GNNs and the linearity of the Koopman operator, the wGKBF model is well suited for accurate prediction of the dynamics of complex networked systems, as illustrated through numerical examples next.
% -----------------------------------------------
\section{Results and Discussion}\label{sec:res}
% -----------------------------------------------
To demonstrate the effectiveness of the proposed approach, we investigate three example nonlinear systems. The first is a lightly-damped double pendulum system with dimension 4, which serves to validate the wLDM framework and study the effects of varying weak form hyperparameters through an ablation study.

The second example addresses the Brusselator dynamics, a classic model problem that can become stiff depending on the choice of parameters and poses significant challenges to numerical solvers. By applying the wLDM to this system and comparing to NODEs, we showcase the proposed framework's capability to tackle stiff systems of ODEs, demonstrating its robustness and efficiency in handling such challenging scenarios.

In the third example, we focus on an electrified aircraft system and employ the wGKBF model with control inputs. This example showcases the model's capability in a realistic, multidisciplinary graph-based system, where the interactions between various components are crucial to capture for accurate modeling and control. The wGKBF effectively captures the complex dynamics of the electrified aircraft energy system with a relatively simple bilinear dynamics model, which may be amenable to the design of accurate yet fast controllers in this application domain.

\subsection{Double Pendulum}\label{sec:dp}
\subsubsection{Problem Setup}

In the first numerical example, we consider a damped and controlled double pendulum problem to show the feasibility of the proposed algorithm. The double pendulum is illustrated in Fig. \ref{dp_figure}, and described as:
\begin{align}
    \ddot{\theta}_1 &= (M_2L_1 \dot{\theta}_1^2 \sin\delta\cos\delta + M_2g\sin\theta_2\cos\delta + M_2L_2\dot{\theta}_2^2\sin\delta - \bar{M}g\sin\theta_1 + u_1)/(L_1\rho) - \dot{\theta}_1 \\
    \ddot{\theta}_2 &= (-M_2L_2 \dot{\theta}_2^2 \sin\delta\cos\delta + \bar{M}g\sin\theta_1\cos\delta - \bar{M}L_1\dot{\theta}_1^2\sin\delta - \bar{M}g\sin\theta_1 + u_2)/(L_2\rho) - \dot{\theta}_2
\end{align}
where $\delta=\theta_2-\theta_1$, $\bar{M}=M_1+M_2$, and $\rho=\bar{M}-M_2\cos^2\delta$.  The masses and lengths of the two pendulums are $M_1 = 1 \text{ kg}$ (upper), $M_2 = 1 \text{ kg}$ (lower), $L_1 = 1 \text{ m}$, and $L_2 = 1 \text{ m}$.  Damping terms are added to both pendulums so as to create a stable isolated equilibrium point in the system.  To generate the dataset for training and testing, we simulate the double pendulum system with initial conditions randomly selected for $\theta_i \in [-10^\circ,10^\circ]$ and $\dot{\theta}_i \in [-10^\circ/s,10^\circ/s]$ and control $u_i\in [-0.25, 0.25]\textrm{ N}$. In this example, the control is held constant for each trajectory to reduce the variability in the subsequent error analysis. The more practical case of time-varying control is considered in the third example. A total of 320 trajectories are generated for model training, followed by another 100 trajectories for testing. Each trajectory is simulated with a 4th order Runge-Kutta method for 20s with a step size of 0.01 seconds. All trajectories were normalized to $[0, 1]$.

\insertfig{dp_figure}{0.4}{Illustration of the double pendulum.}

To evaluate the effectiveness of the proposed weak form, we compare the performance of the wLDM to two baseline methods: Long Short-Term Memory (LSTM) networks \cite{Hochreiter1997,Sutskever2014} and Neural Ordinary Differential Equations (NODEs) \cite{Chen2018neural}. The LSTM network is a type of Recurrrent Neural Network, with its ability to capture long-term dependencies and sequences making it particularly suited for sequential data analysis. As a discrete-time model, LSTM networks %process the data in an Euler fashion, making predictions of 
predict system observations at fixed time steps. In this case, a standard LSTM network implementation processes $L$ steps of observation and control $(\vY,\vU)$ through two LSTM layers, followed by a linear output layer that predicts the observation at the next step $\tilde{\vy}$. On the other hand, NODEs are chosen as a continuous-time baseline model. Including NODEs as a baseline allows for a direct comparison between the proposed wLDM and a state-of-the-art continuous-time approach. The NODE is implemented with multiple layers of FCNN and PReLU activation functions, which integrates a predictive trajectory based on the initial condition and control signals and is then trained with the procedures outlined in Sec. \ref{sec:indirect}. 

The hyperparameters of the nominal wLDM, with a total of 1,481 trainable parameters, are outlined in Table \ref{tbl:wLDM} and follow the architecture specified in Sec. \ref{sec:wldm}.  The nominal model serves as the basis of comparison for the subsequent benchmark against different methods and wLDM's having different parameters.  To ensure a fair comparison, both baseline methods are implemented to match the size of the nominal wLDM. The LSTM comprises a single LSTM layer and an output layer of FCNN (both with dimension $S=32$) followed by PReLU activation, totaling 5,191 parameters. The NODE shares the same architecture as the nominal wLDM, as outlined in Table \ref{tbl:wLDM}, resulting in an identical parameter count.
% Both baseline models are implemented to match the size of the wLDM, ensuring a fair comparison by maintaining a similar parameter count across all models. \todo{Yin - What are the architecture and number of parameters for LSTM and NODE?  Use 1-2 concise sentences to describe the models.} Each model was trained using the Adam optimizer \cite{Diederik2017} with an exponentially decaying learning rate.
% for a total of 5000 iterations to ensure convergence. 

Throughout this and the following examples, all models are implemented using the open-source machine learning framework, PyTorch \cite{PyTorch2019}.  The training and assessment of all the models are performed using a single NVIDIA RTX3090 GPU.

\begin{table}[htbp]
\centering
\caption{Detailed hyperparameters of the nominal wLDM.}
\begin{tabular}{@{}lll@{}}
\toprule
\textbf{Component} & \textbf{Layer Type} & \textbf{Parameters} \\ \midrule
\textbf{Encoder}   & FCNN       & $N_E=1$, $S=32$ \\
          & Activation & PReLU     \\ \midrule
\textbf{Processor} & FCNN & $N_P=1$, $S=32$  \\
          & Activation & PReLU     \\ \midrule
\textbf{Decoder}   & FCNN       & $N_D=1$, $S=32$ \\
          & Activation & PReLU     \\ \bottomrule
\end{tabular}\label{tbl:wLDM}
\end{table}

\subsubsection{Numerical Results}

To compare the model predictions, we train each model on the dataset generated from the double pendulum system with the specified range of parameters. The trained models are then used to predict the future states of the system given initial conditions and control inputs that are unseen in the training dataset, as outlined in Sec. \ref{sec:wldm}. The predictive accuracy of the models is quantified by Normalized Root Mean Square Error (NRMSE).  For a trajectory of $L$ time steps, the NRMSE is defined as, 
\begin{equation}
    \text{NRMSE}=\frac{1}{L}\sum_{l=1}^{L}\frac{\sqrt{\frac{1}{n_y}\sum_{i=1}^{n_y}(\tilde{y}_{l,i}-y_{l,i})^2}}{\max_i(y_{l,i})-\min_i(y_{l,i})},
\end{equation} 
where $[y_{l,1},\cdots,y_{l,n_y}]$ denotes the $n_y$ elements of the observations at the $l$th step of the true trajectory.

\insertfig{dp_comparison}{1.0}{Comparison of prediction performance of the wLDM and the two baseline methods for the damped double pendulum dynamics with controls.}

Figure \ref{dp_comparison} compares the model performance of the nominal wLDM against the baseline models for a representative test case. The predictions from the wLDM and NODE closely match the ground truth over the entire prediction horizon, capturing both the transient dynamics and equilibrium points accurately. There is virtually no deviation in the prediction of the wLDM and NODE, with the wLDM achieving the lowest NRMSE of $1.15\times10^{-3}$. Predictions from the LSTM network show notable deviation across all 4 states in both transient and steady-state behavior, leading to a higher NRMSE of $2.64\times10^{-2}$. These deviations, particularly pronounced in long-term predictions, are likely exacerbated by cumulative errors. A comparison of the predictive NRMSE for all three models across the entire test dataset, and training time, is presented in Table \ref{tbl:dp}. The wLDM achieves the best results in both mean and standard deviation of NRMSE, demonstrating a modest improvement over the NODE and a substantial advantage over the LSTM network. The training time of the wLDM is significantly lower than the baseline methods. This showcases the exceptional accuracy and training efficiency of the wLDM approach.

\begin{table}[ht]
\centering
\caption{Summary of model performance for the double pendulum.}
\label{tbl:dp}
\begin{tabular}{@{}lccccc@{}}
\toprule
\multirow{2}{*}{\textbf{Model}} & \multicolumn{4}{c}{\textbf{NRMSE}} & \textbf{Training} \\ \cmidrule(lr){2-5}
 & \textbf{Mean} & \textbf{Std.} & \textbf{Min} & \textbf{Max} & \textbf{Time (s)}   \\ \midrule
wLDM   & \textbf{9.41E-4} & \textbf{5.05E-4} & 1.53E-4 & \textbf{2.87E-3} & \textbf{2940}  \\
NODE   & 1.11E-3 & 8.03E-4 & \textbf{1.08E-4} & 4.98E-3 & 25480             \\
LSTM  & 2.28E-2 & 1.26E-3 & 1.96E-2 & 2.64E-2 & 64150  \\ \bottomrule
\end{tabular}
\end{table}

\subsubsection{Parametric Study}
To delve deeper into the characteristics of training with a weak form approach, we performed parametric studies with the wLDM and the double pendulum dynamics, varying the time horizon $N$, the order of Jacobi polynomials PO, and integration order IO.
To ensure the fair comparison among the models of different combinations of $N$, PO and IO, the latent space dimension $S$ is increased from 32 to 128.  This is because as PO increases, the wLDM may learn more complex dynamical behaviors from data and would require a model having more parameters to characterize such increased complexity.
% To ensure the model has sufficient capacity to learn the system dynamics as the PO increases, we increased the latent space dimension $S$ from 32 to 128. This allows the model to accommodate the additional complexity introduced by higher-order polynomials without being limited by the model's size.

\insertfig{dp_ablation_N}{1.0}{Comparison of predictive performance for models trained with increasing time window horizon $N$.} % The left plot shows results from models with $\text{PO}=1$ and $\text{IO}=1$, while the models in the right plot are trained with $\text{PO}=2$ and $\text{IO}=2$.}

As a first parametric study, we fixed the PO and IO to isolate and understand the effect of incrementally increasing the time horizon $N$, i.e., the number of steps in the time interval of a test function, on the model's predictive performance. The results in Fig. \ref{dp_ablation_N} demonstrate the impact of the time horizon $N$ on the predictive performance of the wLDM. As $N$ increases, we observe a consistent decrease in the predictive NRMSE, supporting the intuition discussed in Sec. \ref{sec:choice} that using a longer trajectory segment makes the integral constraints more accurate representations of dynamics and leading to a more accurate model. This improvement in performance is particularly noticeable when $N$ is increased from smaller values (e.g., from 3 to 11), indicating that even a modest increase in the time horizon can lead to significant gains in accuracy.  However, the gains in predictive performance become less significant as $N$ grows larger, evident from the diminishing returns in NRMSE reduction beyond $N=27$ in both plots, suggesting a trade-off between model size and computational efficiency.
Furthermore, comparing the left plot ($\text{PO}=\text{IO}=1$) and the right plot ($\text{PO}=\text{IO}=2$), we observe that the model with higher PO and IO achieves lower NRMSE values for the same $N$. This suggests that higher-order polynomials and integration schemes can help capture the system dynamics more effectively, potentially reducing the required $N$ for a given level of performance.  The findings concerning the time interval $N$ are empirical and worth more in-depth theoretical analysis, which can be a valuable future research direction.

\insertfig{dp_ablation_O}{1.0}{Trained wLDM performance with varying PO and IO at a fixed $N=61$ with $95\%$ confidence intervals computed using a bootstrap method.}

The second parametric study investigates the effects of increasing the PO and IO on the model's predictive performance.  To allow for the testing of higher integration orders, which require more time steps, the time horizon $N$ is chosen to be 61. The results, shown in Fig. \ref{dp_ablation_O}, indicate that higher-order polynomials in the weak form offer more flexibility in representing complex dynamics, and higher-order integration schemes provide better approximations of the integral constraints. As PO increases from 1 to 6, the mean predictive NRMSE decreases consistently across all integration orders, suggesting that with a sufficient time horizon $N$, increasing the expressiveness of the test functions through higher-order polynomials leads to more accurate models.

Moreover, for a given PO, increasing the IO generally results in lower NRMSE values and tighter confidence intervals, indicating that higher-order integration schemes lead to more accurate and confident predictions by better approximating the integral constraints. The confidence intervals are computed using a bootstrap method, where a subset of the RMSE values is randomly sampled with replacement, and the mean RMSE is calculated for each bootstrap sample. The tighter confidence intervals associated with higher IO values demonstrate that the model's predictions are more consistent and reliable when using higher-order integration schemes. However, the gains in performance become less pronounced at higher orders, suggesting a trade-off between the increased computational cost in training (due to more constraints associated with higher-order polynomials) and the resulting performance of the model. The optimal choice of PO and IO should balance the desired level of accuracy with the computational cost associated with higher orders. 

\subsection{Brusselator Dynamics}
\subsubsection{Problem Setup}
In this numerical example, we investigate the Brusselator system, a classic model that originates from chemical kinetics and consists of a pair of nonlinear ODEs \cite{Kim2021stiff}. 
The governing equations of the Brusselator model are:
\begin{align}
\dot{x}_1 &= A + x_1^2x_2 - (B+1)x_1 \\
\dot{x}_2 &= B x_1 - x_1^2x_2
\end{align}
where $A$ and $B$ are constants that determine the stiffness of the system.
This model has tunable numerical stiffness so it is suitable for assessing the capabilities of different methods to learn stiff systems.

To assess the predictive accuracy and training efficiency of the wLDM and NODE in handling stiff systems, we generate training data with $A=1$ while varying $B$ between 2 and 5. The increasing stiffness of the system is quantified by computing the stiffness ratio $S_R$, defined as the ratio of the maximum to the minimum absolute eigenvalues of the Jacobian matrix:
\begin{equation}
S_R = \frac{\max(|\lambda_i|)}{\min(|\lambda_i|)}
\end{equation}
where $\lambda_i$ are the eigenvalues of the Jacobian matrix of the Brusselator system, given by:
\begin{equation}
J = \begin{bmatrix}
2x_1x_2 - (B+1) & x_1^2 \\
B - 2x_1x_2 & -x_1^2
\end{bmatrix}
\end{equation}
The stiffness ratio is computed for each trajectory, confirming the increase in stiffness as $B$ increases, with a minimum of 4.33 at $B=2$ and a maximum of 261.21 at $B=5$.

Initial conditions for the trajectories were randomly chosen from within the range $[0, 2]$ for both $x_1$ and $x_2$. Each trajectory was evolved over a time period of $t \in [0, 20]$ seconds using the backward differentiation formula with a timestep of $\Delta t = 0.2$ s. To adequately train the models, 200 trajectories were created for each value of $B$, complemented by an additional set of 100 trajectories for model testing.

The nominal wLDM used in this example has a latent space dimension of $S=128$ and $N_P=2$ processor layers to ensure sufficient capacity for handling the more challenging stiff dynamics. The model was trained with increasing time horizons, from $N=61$ to $N=241$, to capture both the long-term dynamics and short-term spikes accurately. This approach proved particularly useful in the prediction of highly stiff cases, such as $B=5$; in the case of $B=5$, the system exhibits typical phenomenon in stiff dynamics, where the response is smooth most of the time but sudden large variations appear within a short time window. The PO and IO are both set to be 4.

The baseline NODE model was implemented with a similar architecture, matching the number of layers and hidden dimensions to ensure a fair comparison. The models were trained until the relative change in the loss function fell below a tolerance of $10^{-4}$ or a maximum of 10,000 iterations was reached. % All models were trained on a single NVIDIA RTX3090 GPU, which allows us to compare the training efficiency of the two models.

\subsubsection{Numerical Results}
\begin{figure}[ht]
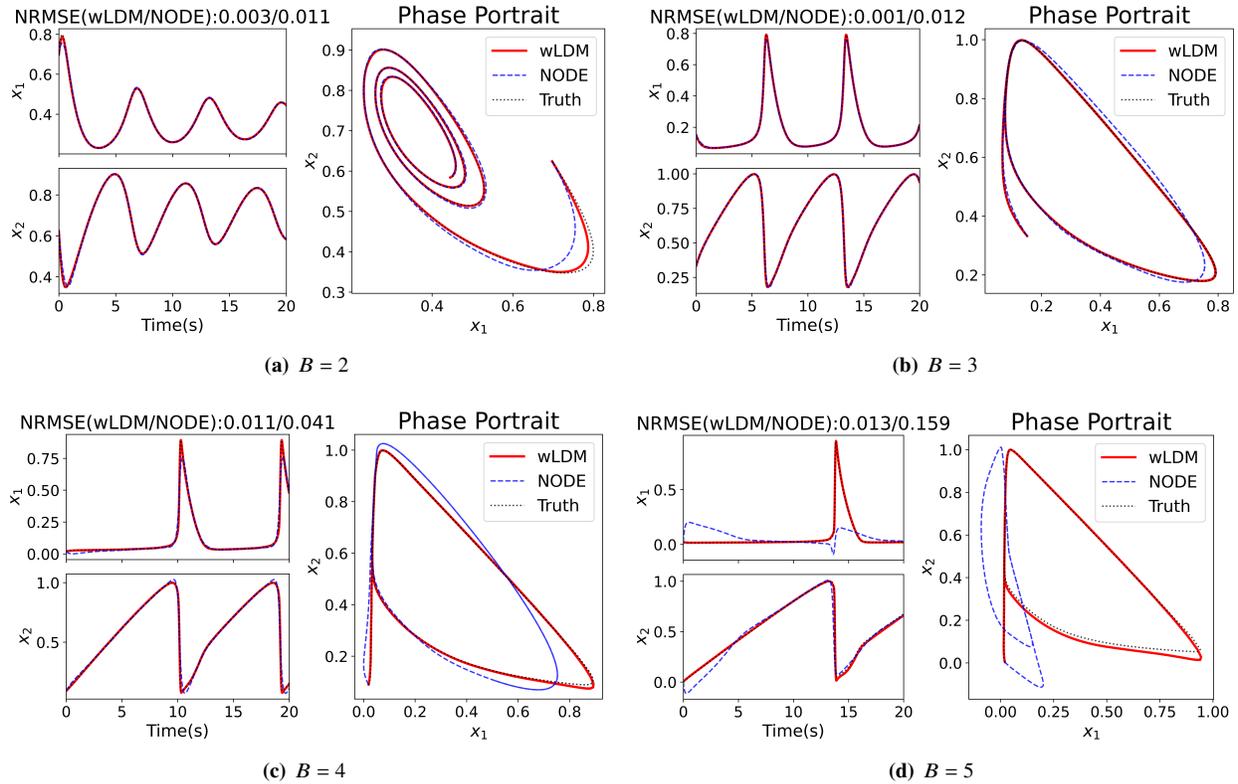

  \centering
  \insertfigs{brusselator_B2}{0.49}{$B=2$}
  \insertfigs{brusselator_B3}{0.49}{$B=3$}
  \insertfigs{brusselator_B4}{0.49}{$B=4$}
  \insertfigs{brusselator_B5}{0.49}{$B=5$}
  \caption{Predictive comparison of the wLDM and baseline NODE model on Brusselator dynamics with increasing stiffness.}
  \label{fig:brusselator}
\end{figure}

Figure \ref{fig:brusselator} compares the predictive accuracy of the wLDM and the baseline NODE model on Brusselator dynamics with increasing stiffness. As $B$ increases, the system undergoes a transition from decaying oscillations converging to a limit cycle ($B=2$) to highly stiff dynamics with sharp, almost discontinuous transitions ($B=5$). These abrupt changes in the system's behavior pose significant challenges to the models in terms of both training and prediction.

The comparison of the NRMSE of the wLDM and NODE are shown in Table \ref{tbl:br}.
The NRMSE values highlight the wLDM's consistently better performance as compared to NODE across all stiffness levels, and the performance gap becomes more pronounced as $B$ increases. For $B=4$, NODE struggles to capture the sharp peaks accurately, resulting in a higher NRMSE. In the highly stiff case of $B=5$, NODE fails to capture the stiffness of the state $x_1$, leading to a substantial increase in the NRMSE.

\begin{table}[ht]
\centering
\caption{Summary of model performance for the Brusselator dynamics.}
\label{tbl:br}
\begin{tabular}{@{}lcccc@{}}
\toprule
\textbf{Model} & $B=2$ & $B=3$ & $B=4$ & $B=5$ \\ \midrule
wLDM   & \textbf{0.003} & \textbf{0.001} & \textbf{0.011} & \textbf{0.013} \\
NODE   & 0.011 & 0.012 & 0.041 & 0.159          \\ \bottomrule
\end{tabular}
\end{table}

The phase portrait analysis in Fig.~\ref{fig:brusselator} provides valuable insights into the models' ability to capture the qualitative behavior of the Brusselator system. In the case of $B=2$, both the wLDM and NODE successfully reproduce the spiral convergence towards a stable limit cycle, indicating their capability to learn the overall dynamics in a less stiff scenario. At $B=3$, the wLDM can accurately capture the elongated and asymmetric shape of the limit cycle, closely matching the ground truth. In contrast, NODE shows deviations, particularly in the outer regions of the limit cycle.

In the highly stiff scenarios ($B=4$ and $B=5$), the advantages of the wLDM become even more pronounced. This model proves more adept at reproducing the complex loops, spirals, and scattered trajectories characteristic of the Brusselator system in these regimes. This can be attributed to the use of higher-order polynomials and integration schemes in the weak form, which provide more flexibility in representing stiff dynamics with rapid transitions. On the other hand, NODE struggles to match the ground truth in these highly stiff cases, especially in the regions of stiff dynamics with rapid transitions. This is likely due to the challenges associated with the numerical integration of stiff ODEs, which can lead to instability and inaccuracies in the learned dynamics. 

\insertfig{brusselator_efficiency}{0.8}{Comparison of training efficiency.}

Figure \ref{brusselator_efficiency} compares the training efficiency of the wLDM and NODE models by plotting the training time per iteration against the number of iterations required for convergence. The results demonstrate that the wLDM requires significantly less computational resources, with each iteration taking less than 1s and convergence achieved in around 4500 iterations. In contrast, NODE models require longer training times per iteration and more iterations to converge, resulting in a much larger total training time as compared to the wLDM. Particularly, the training time per iteration of the NODE models increases as the stiffness increases, but that of the wLDM is insensitive to the stiffness.  While the NODE models require more iterations to converge with increased stiffness, there is an exception of $B=5$, noted specially using a hollow marker in Fig. \ref{brusselator_efficiency}.  In this case, early convergence is achieved but with poor prediction performance; this phenomenon is also observed in other studies \cite{Kim2021stiff}.

Overall, this example demonstrates the superior performance of the wLDM as compared to the baseline NODE model in learning stiff ODEs, consistently achieving better predictive accuracy, capturing intricate details and complex behaviors, and faithfully representing qualitative features in the phase portrait analysis. Moreover, the wLDM exhibits significantly faster training times and converges within a consistent range of iterations, making it a more computationally efficient and practical choice for modeling stiff ODEs.
\subsection{Electrified Aircraft System}\label{sec:aircraft}

\subsubsection{Problem Setup}
\begin{figure}[tb]
\centering
\includegraphics[width=.8\columnwidth]{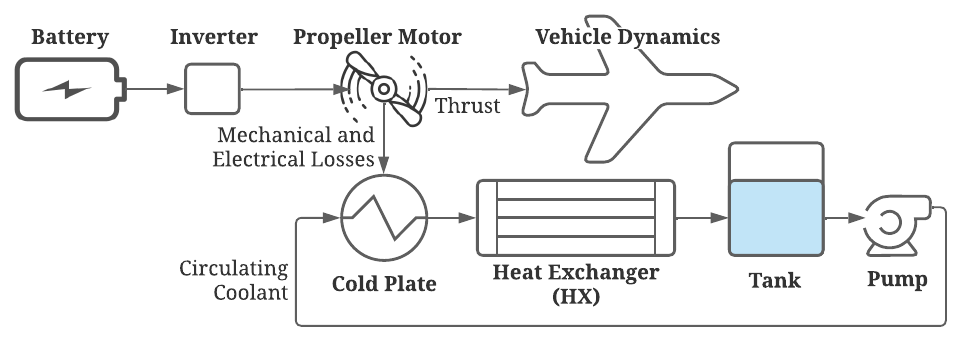}
\caption{Schematic of the electrified aircraft system.}
\label{fig:Schematic of an electrified aircraft system}
\end{figure}

Finally, for a more realistic and complex example, we consider the electrified aircraft system shown in Fig.~\ref{fig:Schematic of an electrified aircraft system}. This electro-thermo-mechanical system consists of a battery delivering power to an inverter and motor, a simplified 1-D model of the longitudinal vehicle dynamics, and a single-phase thermal management system responsible for cooling the motor, consisting of a cold plate, heat exchanger, and tank \cite{Yu2023}. 

\begin{figure}[tb]
\centering
\includegraphics[width=.8\columnwidth]{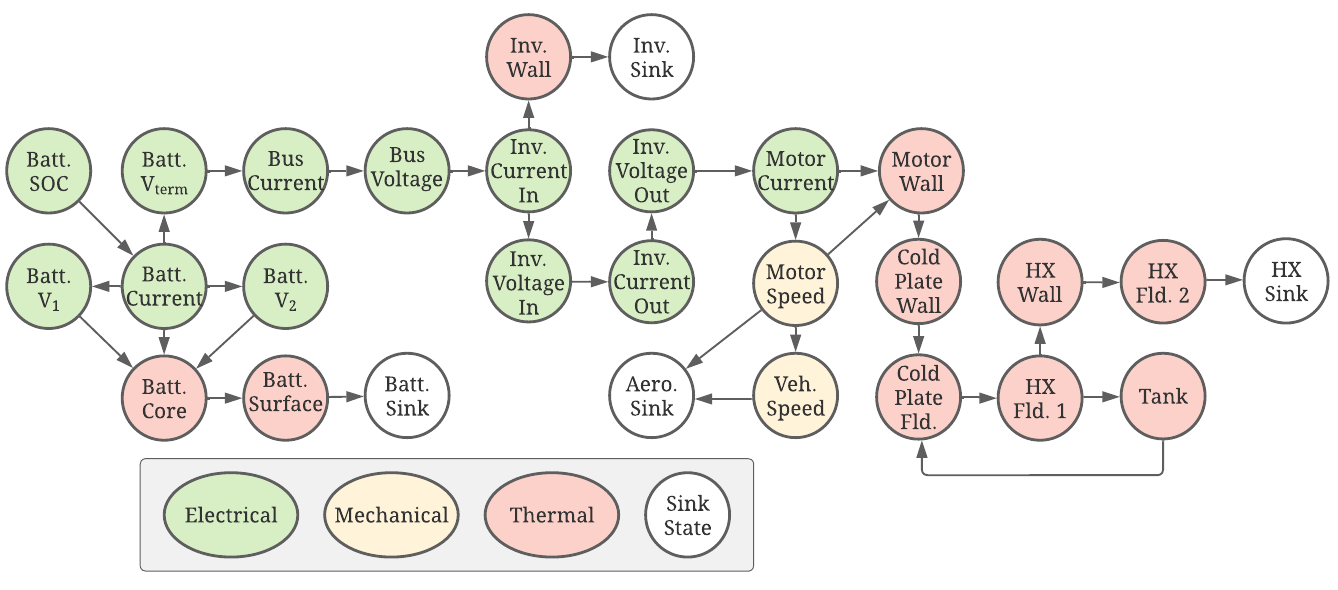}
\caption{Graph-based model structure of the electrified aircraft system.}
\label{fig:Graph-based model of an electrified aircraft system}
\end{figure}

The system dynamics are modeled using the modular graph-based modeling approach from \cite{Aksland2019,Pangborn2018}, producing the oriented graph shown in Fig.~\ref{fig:Graph-based model of an electrified aircraft system} where each node represents a dynamic state or exogenous coupling term and edges indicate which states directly exchange energy. The full-order system, shown in Fig.~\ref{fig:Graph-based model of an electrified aircraft system}, has 24 dynamic states and 3 control inputs, specifically the inverter duty cycle, primary coolant flow rate driven by the pump, and secondary flow rate through the other side of the heat exchanger.

In this example, the physics-based model \cite{Yu2023} is assumed to be the true model that generates the training data, and the goal is to demonstrate the capability of wGKBF model to learn the nonlinear dynamics of the energy system.  A future step is to use experimental measurements of real hardware system to train the wGKBF model.

We apply the wGKBF model as a method of data-driven control-oriented modeling that incorporates the graph structure of the system.
To train the wGKBF model, we generate 75 trajectories by simulating the system with randomly perturbed initial conditions, constrained within the predefined operational boundaries. An additional 25 trajectories are generated for testing the trained model's predictive performance. The system is subjected to chirp control signals with initial and final frequencies randomly sampled from 1 to 4 Hz, ensuring a rich and complex excitation of dynamics across the system. The data are again generated using the backward differentiation formula, with a simulation time of 2 seconds and $\Delta t=0.001$. The resulting time series data is then normalized to the range [0, 1] before being used for model training and testing. 

We benchmark the predictive performance of wGKBF against the conventional implementation of KBF, i.e., the biDMDc approach \cite{Goldschmidt2021}. The biDMDc model utilizes past system states, which is a form of time-delay embedding that incorporates historical data to enhance predictive accuracy. We use 10 steps of past physical states of the system and a truncation order of 20 for the DMD modes, determined after a parametric sweep to achieve the best possible predictive accuracy. To ensure a fair comparison, the wGKBF model also incorporates the same 10 steps of past physical states as node features in the graph. Theoretically, the graph-based KBF could be trained without the weak form, e.g., using the conventional graph NODE methods \cite{Poli2019,Zang2020,Luo2023,Huang2023}.  However, these methods resulted in intractable computational cost in the learning of the electrified aircraft dynamics due to the high numerical stiffness. Hence in this study only the wGKBF and biDMDc results are reported and compared.
% \todo{Yin - Any lifting in biDMDc?}

\begin{table}[htbp]
\centering
\caption{Detailed hyperparameters of the nominal wGKBF model.}
\begin{tabular}{@{}lll@{}}
\toprule
\textbf{Component} & \textbf{Layer Type} & \textbf{Parameters} \\ \midrule
\textbf{Encoder}   & GCL       & $N_E=2$, $n_N=24$, $\tilde{S}=16$ \\
          & Activation & PReLU     \\ \midrule
\textbf{Processor} & KBF ($\vA,\vB_i)$ & $S=384$, $i=1,2,3$  \\\midrule
\textbf{Decoder}   & GCL       & $N_D=2$, $\tilde{S}=16$ \\
          & Activation & PReLU     \\ \bottomrule
\end{tabular}\label{tbl:wGKBF}
\end{table}

The hyperparameters of the nominal wGKBF model are summarized in Table \ref{tbl:wGKBF}.  In this case, the states at all nodes share the same dimension of 1, hence the FCNN steps in the encoder and decoder are removed.  The encoder and decoder both consist of two GCL layers with PReLU activation functions. The latent space dimension is set to $\tilde{S}=16$, determined through a parametric sweep to optimize model prediction accuracy while maintaining a tractable number of parameters in the KBF processor. The number of nodes in the graph, $n_N=24$, matches the number of system states. Consequently, the KBF processor has a state dimension of $S=384$, with three input matrices $\vB_i$ corresponding to the three control inputs.
\insertfig{aircraft_chirp}{0.8}{Comparison of wGKBF and biDMDc prediction for the electrified aircraft system.}

\subsubsection{Numerical Results}

Figure \ref{aircraft_chirp} compares the performance of wGKBF and biDMDc in predicting the simulated aircraft system dynamics with chirp control signals. The true system dynamics, represented by the black dashed lines, exhibit a mix of steady, slowly varying, and rapidly fluctuating behaviors across different state variables. The thermal states, such as cold plate fluid and wall temperatures ($x_1$ and $x_2$), show slow response characteristics, while some electrical states, including inverter voltages and currents ($x_{16}$, $x_{17}$, $x_{19}$, and $x_{20}$), display fast dynamics with significant variations. The mechanical states, like motor speed ($x_{22}$) and vehicle speed ($x_{24}$), demonstrate intermediate dynamics.

The wGKBF model exhibits superior performance in capturing the true dynamics of the electrified aircraft system, achieving an NRMSE of 0.026, significantly lower than the NRMSE of biDMDc of 0.475. The wGKBF's ability to accurately predict the rapid fluctuations in electrical states and the subtle variations in thermal states highlights its effectiveness in handling the multi-domain and multi-timescale dynamics of the electro-thermo-mechanical system. 

This enhanced performance can be attributed to several aspects of the wGKBF model. The graph-based representation allows the wGKBF to explicitly model the interactions and dependencies among the various components of the aircraft system. By incorporating the system's inherent structure into the model architecture, the wGKBF can more effectively capture localized features and propagate information across the graph, leading to a more accurate and interpretable representation of the system dynamics. Furthermore, the deep learning nature of the wGKBF model endows it with a high level of expressiveness, allowing it to learn complex nonlinear mappings between the input and output spaces. The combination of GCLs and KBF in the wGKBF architecture enables the model to capture the dynamics of multiple timescales of the system, resulting in improved predictive accuracy compared to the biDMDc baseline.

% Figure \ref{aircraft_chirp} compares the performance of two models in predicting the simulated aircraft system dynamics with chirp control signals. The true system dynamics, as indicated by the dashed lines in the graph, represent the real-world behavior of an electrified aircraft's electro-thermo-mechanical system. Most of the states exhibit a steady or slowly varying behavior over time, which is particularly evident in the thermal states such as the cold plate fluid and wall temperatures ($x_1$ and $x_2$), reflecting the inherent slow response of thermal processes. In contrast, the electrical states, such as inverter voltages and currents ($x_{16}$,$x_{17}$,$x_{19}$, and $x_{20}$), show a more dynamic behavior with significant variations over time. This is indicative of the fast response characteristics of electrical systems to control inputs and changes in load. The mechanical states like motor speed ($x_{22}$) and vehicle speed ($x_{24}$) also display intermediate dynamics, changing more quickly than the thermal states but not as abruptly as the electrical states.

\insertfig{aircraft_chirp_nrmse_all_testcases}{0.7}{Comparison of predictive performance of wGKBF and biDMDc across all test cases.}

% In comparing the model performances depicted in Fig. \ref{aircraft_chirp}, the wGKBF exhibits a superior fit to the true dynamics of the electric aircraft system, with a NRMSE of 0.026, significantly lower than the biDMDc which stands at an RMSE of 0.475. The wGKBF's accuracy across all state variables—particularly in capturing the rapid fluctuations in the electrical states and the more subtle variations in the thermal states—demonstrates its robust capability to handle the integrated and multi-scaled dynamics of the electro-thermo-mechanical system. Furthermore, the predictive performance of the two models across all 25 testing cases are shown in Fig. \ref{aircraft_chirp_nrmse_all_testcases}, and summarized in Table \ref{tbl:aircraft}. The wGKBF model, shown in red, maintains a consistently low NRMSE across all test cases, indicating a high level of predictive accuracy and consistency. While the performance of biDMDc model, shown in represented by the blue dashed line, shows greater variability in prediction error, with NRMSE values generally higher than those of the wGKBF. The averaged NRMSE The stark contrast between the two models highlights the enhanced accuracy and stability of the wGKBF model in predicting the dynamics of the aircraft system over the tested  scenarios.

\begin{table}[ht]
\centering
\caption{Summary of model performance for the electrified aircraft system.}
\label{tbl:aircraft}
\begin{tabular}{@{}lccccc@{}}
\toprule
\multirow{2}{*}{\textbf{Model}} & \multicolumn{4}{c}{\textbf{NRMSE}} & \textbf{Training} \\ \cmidrule(lr){2-5}
 & \textbf{Mean} & \textbf{Std.} & \textbf{Min} & \textbf{Max} & \textbf{Time (s)}   \\ \midrule
wGKBF   & \textbf{3.38E-2} & \textbf{1.12E-2} & \textbf{6.47E-2} & \textbf{1.84E-2}&  16680    \\
biDMDc   & 5.82E-1 & 6.23E-2 & 6.97E-1 & 4.73E-1 & \textbf{724}   \\\bottomrule
\end{tabular}
\end{table}

The predictive performance of the two models across all 25 test cases is shown in Fig. \ref{aircraft_chirp_nrmse_all_testcases}, and summarized in Table \ref{tbl:aircraft}. The wGKBF maintains a consistently low NRMSE with a mean value of $3.38\times 10^{-2}$ and a small standard deviation of $1.12\times 10^{-2}$.
% , demonstrating its robustness and reliability in predicting the aircraft system dynamics under various scenarios.
In contrast, the biDMDc model, while takes significantly less time to train,
% , which lacks the graph-based representation and deep learning architecture, 
shows greater variability in prediction error, with higher NRMSE values and a larger standard deviation.
The substantial gap in performance between the two models highlights the effectiveness of the wGKBF approach in leveraging the graph topology and deep learning to capture the complex nonlinear dynamics of the electrified aircraft system. By incorporating the system's structure and learning expressive mappings, the wGKBF model achieves significantly improved predictive accuracy and consistency, showcasing its superior capability in modeling the complex electro-thermo-mechanical systems.

% -----------------------------------------------
\section{Conclusions}\label{sec:con}
% -----------------------------------------------

Accurate and computationally efficient control-oriented models are essential for the design and implementation of advanced control strategies to optimize system performance, safety, and efficiency of networked systems, such as electrified aircraft energy systems.
The development of such models requires answering two questions: What model form to use and how to train the model?
This paper addresses the second question by introducing a novel deep learning approach, the weak Latent Dynamics Model (wLDM), for learning generic nonlinear dynamics with control.  Leveraging the weak form, the wLDM enables more numerically stable and computationally efficient training as well as more accurate predictive model, when compared to traditional approaches. Addressing the first question, we build upon the wLDM and propose the weak Graph Koopman Bilinear Form (wGKBF) model that integrates graph-based representations and Koopman theory to learn control-oriented bilinear dynamics for networked systems.

The effectiveness of the wLDM and its wGKBF variant is demonstrated through three example systems of increasing complexity: 
\begin{compactenum}
\item A controlled double pendulum: The numerical experiments show that the wLDM consistently outperforms baseline models in terms of predictive accuracy and training efficiency. Parametric studies provide valuable insights into the role of hyperparameters in the weak form. The time horizon $N$ is found to have a significant impact on predictive performance, with larger values leading to more accurate models at the cost of increased training time. The order of Jacobi polynomials (PO) and integration order (IO) also play crucial roles, with higher orders capturing more complex dynamics but exhibiting diminishing returns in performance gains.
\item The Brusselator dynamics: A systematic study is performed to demonstrate the capability of the wLDM to learn stiff dynamics, which are challenging for conventional learning algorithms such as NODEs.  It is found that the wLDM outperforms the NODEs in both predictive accuracy and training efficiency.  In particular, the NODEs take substantially longer time to train as the stiffness increases while the training time of the wLDM is shown to be insensitive to the stiffness.  This comparison highlights the potential of the wLDM for learning complex stiff dynamics, such as the systems having disparate timescales.
\item An electrified aircraft energy system: The wLDM is extended to wGKBF by incorporating graph-based representations and the Koopman bilinear form. The wGKBF model effectively captures the intricate couplings and dependencies among the various components of the electrified aircraft energy system and outperforms the existing bilinear system identification method in predictive accuracy by an order of magnitude. % These advancements open up new possibilities for modeling and control of complex, real-world systems.
\end{compactenum}

Collectively, the wLDM and wGKBF model offer a powerful and flexible framework for learning the dynamics of complex networked systems in aerospace applications. By leveraging the weak form, graph-based representations, and Koopman theory, these models enable accurate, efficient, and control-oriented modeling of challenging systems, opening up new possibilities for advanced system modeling and control. Future work will focus on leveraging these models for the design of advanced control strategies, such as hierarchical MPC and reinforcement learning-based approaches, as well as system analysis tasks, including stability assessment, sensitivity studies, and optimization of system configurations. Additionally, investigating the scalability and generalizability of the proposed framework to larger-scale systems with higher-dimensional state spaces and more intricate network structures will be a key direction for further research.

\appendix
\section{Koopman Theory and Bilinear Form}\label{app:kbf}

Koopman theory provides a convenient tool for the analysis and control of nonlinear dynamical systems. When the system is autonomous and has an isolated stable equilibrium point, a global linearization can be achieved in the entire attracting basin of the system \cite{Budisic2012}.  With such linearization, classical linear system theory can be leveraged to characterize the nonlinear dynamics. For a controlled system, the Koopman operator can be extended to the bilinear form \cite{Goswami2022}, allowing for the efficient representation and control of nonlinear systems with inputs. For the more general controlled case, see Ref. \cite{Proctor2018}.

Consider a control-affine system of the form
\begin{equation}\label{eqn:controlAffine}
    \dot{\vx}=\vf_0(\vx)+\sum_{i=1}^{n_u}\vf_i(\vx)u_i\:,\quad \vx(0)=\vx_0\:,
\end{equation}
where $\vx\in\bX\subseteq\bR^{n_x}$ and $\vu=[u_1\dots u_{n_u}]^T\in\bR^{n_u}$ are states and control, respectively, $\vf_0:\bX\rightarrow\bX$ captures the autonomous dynamics, and $\vf_i:\bX\rightarrow\bX$ captures the influence of control actions on the system.

In the absence of control inputs \cite{Mauroy2016}, i.e., when $\vu=0$, the system generates a flow $\vF_t(\vx_0)=\vx(t)$ from an initial condition $\vx_0$.  The continuous time Koopman operator $\koop:\cF\rightarrow\cF$ is an infinite-dimensional linear operator such that $\koop z = z\circ\vF_t$ for all $z\in\cF$, where $z:\bX\rightarrow\bC$ is a complex-valued observable function of the state vector $\vx$, $\cF$ is the function space of all possible observables, and $\circ$ denotes function composition.
As a linear operator, $\koop$ admits eigenpairs $(\lambda,\varphi)$ such that
\begin{equation}\label{eqn:koopman-eigen}
    \koop\varphi = \varphi\circ\vF_t = e^{\lambda t}\varphi\:,
\end{equation}
where $\lambda\in\bC$ and $\varphi\in\cF$ are the Koopman eigenvalue and Koopman eigenfunction, respectively.

The infinitesimal generator of $\koop$ associated with $\vf_0$, referred to as the Koopman generator, is defined as $\cL_{\vf_0}=\lim_{t\rightarrow0}\frac{\koop-I}{t}$, where $I$ is the identity operator, and turns out to be the Lie derivative $\cL_{\vf_0}=\vf_0\cdot\nabla$, with eigenpair $(\lambda,\varphi)$,
\begin{equation}
    \dot{\varphi} = \cL_{\vf_0}\varphi = \lambda\varphi\:.
\end{equation}

Given a set of eigenpairs $\{(\lambda_i,\varphi_i)\}_{i=1}^n$, the {Koopman Canonical Transform} (KCT) \cite{Surana2016} of the control-affine system in Eq. \eqref{eqn:controlAffine} is given as 
\begin{equation}
    \dot{\vtvf} = \vtL\vtvf + \sum_{i=1}^{n_u}\cL_{\vf_i}\vtvf u_i\:,
\end{equation}
where $\vtL=\diag([\lambda_1,\cdots,\lambda_n])$, $\vtvf=[\varphi_1,\cdots,\varphi_n]$, and Lie derivatives for the control terms are $\cL_{\vf_i}=\vf_i\cdot\nabla$.

Suppose the set of eigenfunctions is sufficiently large, such that $\vtvf$ span an invariant space for $\cL_{\vf_i}$, \ie, each $\cL_{\vf_i}$ can be represented using an $n\times n$ 
matrix $\vD_i$ such that $\cL_{\vf_i}\vtvf=\vD_i\vtvf$. Then the KCT can be brought to a bilinear form \cite{Goswami2022},
\begin{equation}\label{eqn:KCT-bi}
    \dot{\vtvf} = \vtL\vtvf + \sum_{i=1}^{n_u}\vD_i\vtvf u_i\:.
\end{equation}
Often it is difficult to directly obtain the eigenfunctions of $\koop$. Instead, by introducing lifted coordinates via a mapping $\vw=\vtf(\vx)$ such that $\vtvf=\vtY^H\vw$, and substituting it into the KCT in bilinear form from Eq. (\ref{eqn:KCT-bi}), we reach the commonly used Koopman Bilinear Form (KBF) \cite{Goswami2022,Jiang2024},
\begin{equation}\label{eqn:kbf}
    \dot{\vw}=\vA\vw+\sum_{i=1}^{n_u}\vB_i\vw u_i\:,
\end{equation}
%\vtF^H\dot{\vtvf}=
where an eigendecomposition $\vA=\vtF\vtL\vtY^H$ reproduces the Koopman eigenvalues $\vtL$ and $\vB_i=\vtF\vD_i\vtY^H$. The original states are recovered from an inverse mapping as $\vx=\vtf^{-1}(\vw)\equiv\vty(\vw)$.

The KBF provides a global bilinearization of the control-affine system in Eq. \eqref{eqn:controlAffine}, enabling the application of linear and bilinear control techniques to the nonlinear aircraft energy system. By incorporating the KBF into the wLDM framework, we aim to learn a control-oriented representation of the aircraft energy system that captures its nonlinear dynamics while maintaining a structure amenable to efficient control design.

The condition for bilinearization into the form Eq. (\ref{eqn:kbf}) is restated as follows:
\begin{theorem} \cite{Goswami2022}
Suppose a set of the Koopman eigenfunctions $\{\varphi_1,\varphi_2,\dots,\varphi_n\}$ of the autonomous system forms a invariant subspace of $\cL_{\vf_i},i=1,\dots,m$. Then Eq. (\ref{eqn:controlAffine}) is bilinearizable with an $n$ dimensional state space.
\end{theorem}

\section*{Acknowledgements}
This work was partially supported by a  seed grant from The Pennsylvania State University Institute for Computational and Data Sciences (ICDS) . The research of YY and DH was partially supported by National Science Foundation under the award DMS-2229435.
% \section*{Conflict of Interests}
% The authors declare that they have no known competing financial interests or personal relationships that could have appeared to influence the work reported in this paper.
\bibliography{references}
\end{document}